%% file: main.tex
\documentclass{article}
\usepackage{ijcai23}

\usepackage{times}
\usepackage{soul}
\usepackage{url}
\usepackage[hidelinks]{hyperref}
\usepackage[utf8]{inputenc}
\usepackage[small]{caption}
\usepackage{graphicx}
\usepackage{amsmath}
\usepackage{amsthm}
\usepackage{booktabs}
\usepackage{algorithm}
\usepackage{algorithmic}
\usepackage[switch]{lineno}

\newtheorem{example}{Example}
\newtheorem{theorem}{Theorem}
\newtheorem{definition}{Definition}
\newtheorem{lemma}{Lemma}
\newtheorem{corollary}{Corollary}

\title{Constructive Interpolation and Concept-Based Beth Definability for Description Logics via Sequents}

\author{
Tim S. Lyon
\and
Jonas Karge
\affiliations
Institute of Artificial Intelligence, Technische Universit\"at Dresden\
\emails
\{timothy\_stephen.lyon, jonas.karge\}@tu-dresden.de
}

\input{notation.tex}

\begin{document}

\maketitle

\begin{abstract}
We introduce a constructive method applicable to a large number of description logics (DLs) for establishing the concept-based Beth definability property (CBP) based on sequent systems. Using the highly expressive DL $\riq$ as a case study, we introduce novel sequent calculi for $\riq$-ontologies and show how certain interpolants can be computed from sequent calculus proofs, which permit the extraction of explicit definitions of implicitly definable concepts. To the best of our knowledge, this is the first sequent-based approach to computing interpolants and definitions within the context of DLs, as well as the first proof that $\riq$ enjoys the CBP. Moreover, due to the modularity of our sequent systems, our results hold for restrictions of $\riq$, and are applicable to other DLs by suitable modifications.
\end{abstract}

\section{Introduction}\label{sec:introduction}

\input{introduction.tex}

\section{Preliminaries}\label{sec:log-prelims}

\input{body-1.tex}

\section{Sequent Systems}\label{sec:sequent-systems}

\input{body-2.tex}

\section{Constructive Sequent-Based Method}\label{sec:interpolants}

\input{body-3.tex}

\section{Concluding Remarks}\label{sec:conclusion}

\input{conclusion.tex}

\appendix

\section*{Acknowledgments}
Tim S. Lyon is supported by the European Research Council (ERC) Consolidator Grant 771779 (DeciGUT). Jonas Karge is supported by BMBF (Federal Ministry of Education and Research)
in DAAD project 57616814 (\href{https://secai.org/}{SECAI, School of Embedded Composite AI}) as well as by BMBF in the Center for Scalable Data Analytics and Artificial Intelligence (ScaDS.AI).

\bibliographystyle{named}
\bibliography{bibliography}

\clearpage
\onecolumn

\section{Proofs for \sect~\ref{sec:log-prelims}}

\input{app-0.tex}

\section{Proofs for \sect~\ref{sec:sequent-systems}}\label{app:proofs}

\input{app-1.tex}

\section{Proofs for \sect~\ref{sec:interpolants}}

\input{app-2.tex}

\end{document}

%% file: notation.tex
\usepackage{amssymb, amsmath, mathrsfs, mathtools}
\usepackage{synttree, bussproofs,scrextend, stmaryrd, rotating, xcolor, upgreek}
\usepackage{esvect,pgf, tikz, color}
\usetikzlibrary{arrows, automata, positioning}
\usepackage[all]{xy}
\usepackage{changepage,enumerate,proof,upgreek,relsize,trfsigns,
comment}
\usepackage{float}

\definecolor{tim}{RGB}{0, 0, 250} 

\definecolor{jonas}{RGB}{0, 125, 0} 

\newtheorem{innercustomlem}{Lemma}
\newenvironment{customlem}[1]
  {\renewcommand\theinnercustomlem{#1}\innercustomlem}
  {\endinnercustomlem}
  
  \newtheorem{innercustomthm}{Theorem}
\newenvironment{customthm}[1]
  {\renewcommand\theinnercustomthm{#1}\innercustomthm}
  {\endinnercustomthm}

\newcommand{\iffi}{\textit{iff} }

\newcommand{\sect}{Section}
\newcommand{\fig}{Figure}

\newcommand{\thm}{Theorem}
\newcommand{\cor}{Corollary}

\newcommand{\dfn}{Definition}
\newcommand{\lem}{Lemma}

\newcommand{\calc}{\mathsf{S}(\ont)}
\newcommand{\icalc}{\mathsf{SI}(\ont)}

\newcommand{\cnames}{\mathsf{N_{C}}}
\newcommand{\rnames}{\mathsf{N_{R}}}
\newcommand{\roles}{\mathbf{R}}

\newcommand{\tbox}{\mathcal{T}}
\newcommand{\ont}{\mathcal{O}}

\newcommand{\rbox}{\mathcal{R}}
\newcommand{\lab}{\mathrm{Lab}}

\newcommand{\omodels}{\vDash_{\ont}}

\newcommand{\inv}[1]{\mathsf{Inv}(#1)}
\newcommand{\invn}[1]{{#1}^{-}}

\newcommand{\vocab}{\mathcal{V}}
\newcommand{\card}[1]{|#1|} 

\newcommand{\sig}[1]{\mathrm{sig}(#1)}
\newcommand{\sigtr}{\Theta}
\newcommand{\sigtrii}{\Xi}
\newcommand{\cpt}[1]{\mathrm{con}(#1)}
\newcommand{\role}{r} 
\newcommand{\roleb}{s} 

\newcommand{\alc}{\mathcal{ALC}}
\newcommand{\alcq}{\mathcal{ALCQ}}

\newcommand{\h}{\mathcal{H}}

\newcommand{\nom}{\mathcal{O}}
\newcommand{\ir}{\mathcal{I}}

\newcommand{\negnnf}[1]{\dot{\neg} #1}
\newcommand{\riq}{\mathcal{RIQ}}

\newcommand{\gram}{\mathrm{G}}
\newcommand{\sto}[1]{\gram(#1)}
\newcommand{\albet}{\roles} 

\newcommand{\cate}{ }

\newcommand{\concat}{ }

\newcommand{\glang}{L_{\sto{\ont}}}
\newcommand{\pto}{\longrightarrow}

\newcommand{\dr}{\pto^{*}_{\sto{\ont}}}

\newcommand{\albetstr}{\roles^{*}}
\newcommand{\stra}{S}
\newcommand{\strb}{R}

\newcommand{\strc}{T}
\newcommand{\empstr}{\varepsilon}

\newcommand{\itp}{\mathcal{G}}
\newcommand{\sep}{ \ \| \ }
\newcommand{\splt}{ \ {}^{a}|^{b} \ } 
\newcommand{\spltrev}{ \ {}^{b}|^{a} \ }
\newcommand{\splti}{ \ {}^{1}|^{2} \ }

\newcommand{\orth}[1]{\overline{#1}}

\newcommand{\inter}{\mathcal{I}}
\newcommand{\interii}{\mathcal{J}}
\newcommand{\dom}{\Delta^{\inter}}
\newcommand{\domii}{\Delta^{\interii}}
\newcommand{\map}[1]{{#1}^{\inter}}
\newcommand{\lmap}{\lambda}

\newcommand{\rel}{\Gamma} 
\newcommand{\cxti}{\Delta}
\newcommand{\cxtii}{\Sigma}
\newcommand{\cxtiii}{\Pi}
\newcommand{\cxtiv}{\Pi}
\newcommand{\la}{x}
\newcommand{\lb}{y}
\newcommand{\lc}{z}
\newcommand{\prf}{\pi}
\newcommand{\branch}{\mathcal{B}}

\newcommand{\size}[1]{s(#1)}

\newcommand{\wght}[1]{w(#1)}

\newcommand{\sar}{\vdash}

\newcommand{\seq}{S}
\newcommand{\prgr}[1]{\mathrm{PG}(#1)}

\newcommand{\prgrdom}{V}
\newcommand{\prgredges}{E}

\newcommand{\con}{\sqcap}
\newcommand{\dis}{\sqcup}
\newcommand{\imp}{\sqsubseteq}
\newcommand{\all}{\forall \role.}
\newcommand{\some}{\exists \role.}
\newcommand{\ltn}{{\leqslant} n \roleb.}
\newcommand{\gtn}{{\geqslant} n \roleb.}

\newcommand{\id}{(id)}
\newcommand{\idc}{(id)}
\newcommand{\idcIi}{(id^{I}_{1})}
\newcommand{\idcIii}{(id^{I}_{2})}
\newcommand{\ideq}{(id_{\dot{=}})}
\newcommand{\ideqI}{(id_{\dot{=}}^{I})}

\newcommand{\topr}{(\top)}

\newcommand{\conr}{(\con)}
\newcommand{\conrI}{(\con^{I})}

\newcommand{\disr}{(\dis)}
\newcommand{\disrI}{(\dis^{I})}

\newcommand{\allr}{(\forall \role)}
\newcommand{\allrI}{(\forall \role^{I})}

\newcommand{\existsr}{(\exists \role)}
\newcommand{\existsrI}{(\exists \role^{I})}
\newcommand{\ltnr}{({\leqslant} n \role)}
\newcommand{\ltnrI}{({\leqslant} n \role^{I})}
\newcommand{\gtnr}{({\geqslant} n \role)}
\newcommand{\gtnrI}{({\geqslant} n \role^{I})}
\newcommand{\orthru}{(O)}

\newcommand{\wk}{(w)}
\newcommand{\wkeq}{(w_{\eq})}
\newcommand{\wkneq}{(w_{\noteq})}

\newcommand{\ctr}{(c)}

\newcommand{\ctrr}{(c_{r})}

\newcommand{\prove}{\mathtt{Prove}}
\newcommand{\true}{\mathtt{True}}

\newcommand{\simplepath}[1]{\overset{\mathclap{#1}}{\leadsto}}
\newcommand{\prpath}[1]{\hspace{3pt}\raise-3pt\hbox{$\simplepath{#1}$}\hspace{3pt}}
\newcommand{\gcilist}[1]{\negnnf{\mathcal{T}}_{#1}}
\newcommand{\amodels}{\vDash^{\forall}}
\newcommand{\emodels}{\vDash^{\exists}}
\newcommand{\neqset}[1]{\rel^{\not\dot{=}}(#1)}
\newcommand{\neqsetii}[1]{\rel^{\not\dot{=}}_{i,\ell}(#1)}
\newcommand{\noteq}{\,{\not}{\dot{=}}\,}
\newcommand{\eq}{\,\dot{=}\,}
\newcommand{\subeq}{(s_{\dot{=}})}
\newcommand{\subeqI}{(s_{\dot{=}}^{I})}
\newcommand{\eqpath}[2]{#1 =^{*}_{\rel} #2}
\newcommand{\eqclass}[2]{[#1]_{#2}}
\newcommand{\geqnir}{{\geqslant}{(n{+}1)} r}

\newcommand{\leqnr}{{\leqslant}{n} r}
\newcommand{\clist}{\vv{C}}
\newcommand{\ru}{(r)}
\newcommand{\ruI}{(r^{I})}
\newcommand{\dl}{\mathcal{L}}
\newcommand{\deslog}{\mathcal{L}}
\newcommand{\relab}[2]{(\ell^{#1}_{#2})}
\newcommand{\symneq}{(s_{\noteq})}

\newcommand{\pspace}{\mathsf{PSPACE}}
\newcommand{\exptime}{\mathsf{EXPTIME}}

%% file: introduction.tex
Defining new concepts in terms of given concepts and relations is an important operation within the context of description logics (DLs), and logic more generally. Typically, a new concept $\mathsf{NewC}$ can be defined in one of two ways: (1) \emph{implicitly}, by specifying a set of axioms such that the interpretation of $\mathsf{NewC}$ is uniquely determined by the interpretation of the given concepts and relations, or (2) \emph{explicitly}, by writing a definition $\mathsf{NewC} \equiv \mathsf{D}$ where $\mathsf{NewC}$ does not appear in $\mathsf{D}$. Description logics for which implicit definability implies explicit definability are said to be \emph{definitorially complete}~\cite{BaaNut03,CatConMarVen06}, or to exhibit the \emph{concept-based Beth definability property (CBP)}~\cite{TenFraSey13}. This is Beth's definability property~\cite{Bet56} from first-order logic rephrased for DLs. 

Beth definability and variations thereof have found numerous applications in DLs. For example, the property has been used in ontology engineering to extract acyclic terminologies from general ones~\cite{BaaNut03,CatConMarVen06}. This is of particular importance since reasoning with acyclic terminologies is usually less complex than with general ones, e.g. satisfiability over acyclic $\alc$-terminologies is $\pspace$-complete while being $\exptime$-complete over general $\alc$-terminologies~\cite{Don03}. 
Other applications include, rewriting ontology-mediated queries~\cite{FraKer19,SeyFraBru09,TomWed22}, learning concepts separating positive and negative examples~\cite{ArtJunMazOzaWol23,FunJunLutPulWol19}, and computing referring expressions, which is of value in computational linguistics and data management~\cite{AreKolStr08,BorTomWed16,ArtMazOzaWol21}.

A number of methods have been used to confirm the existence of, or actually compute, explicit definitions of implicitly definable concepts for expressive DLs; e.g. model-theoretic mosaic-based methods have been employed to decide the existence of explicit definitions for $\alc\h$, $\alc\nom$, and $\alc\h\nom\ir$~\cite{ArtJunMazOzaWol23,JunMazWol22}. However, as noted in these works, these methods are \emph{non-constructive}, confirming the existence of explicit definitions without necessarily providing them. Thus, interest has been expressed in developing constructive methods that \emph{actually compute} explicit definitions. We note that constructive methods have been employed in the literature, e.g. methods relying on the computation of normal forms and uniform interpolants~\cite{CatConMarVen06} or which compute explicit definitions using tableau-based algorithms~\cite{TenFraSey13}. With the aim of furthering this programme, we present a constructive method applicable to a large number of DLs, which computes explicit concept-based definitions of implicitly definable concepts and establishes the CBP by means of \emph{sequent systems}.

Since its introduction in the 1930's, Gentzen's sequent calculus has become one of the preferred formalisms for the construction of proof calculi~\cite{Gen35a,Gen35b}. A sequent calculus is a set of inference rules operating over expressions (called \emph{sequents}) of the form $\Gamma \sar \Delta$ with $\Gamma$ and $\Delta$ sequences or (multi)sets of formulae. Sequent systems have found fruitful applications, being exploited in the development of automated reasoning methods~\cite{Sla97} and being used to establish non-trivial properties of logics such as consistency~\cite{Gen35a,Gen35b}, decidability~\cite{Dyc92}, and interpolation~\cite{Mae60}. Regarding this last point, it was first shown by Maehara that sequent systems could be leveraged to \emph{constructively} prove the Craig interpolation property~\cite{Cra57} of a logic. Since this seminal work, Maehara's interpolation method has been extended and adapted in a variety of ways to 
prove Craig interpolation for diverse classes of logics with sequent-style systems, including modal logics~\cite{FitKuz15}, intermediate logics~\cite{KuzLel18}, and temporal logics~\cite{LyoTiuGorClo20}. As Craig interpolation implies Beth definability, it 
 follows that the sequent-based methodology is applicable to the latter.


In this paper, we provide the first sequent calculi for $\riq$-ontologies and show how these calculi can be used to compute interpolants, explicit definitions, and to confirm the CBP. Although our work is inspired by Maehara's method, we note that it is a non-trivial generalization of that method. As discussed in~\cite{LyoTiuGorClo20}, Maehara's original method is quite restricted, being inapplicable in many cases to even basic modal logics, which \emph{a fortiori} means the method is inapplicable to expressive DLs. To overcome these difficulties, we use a generalized notion of sequent and interpolant that encodes a tree whose nodes are multisets of DL concepts accompanied by (in)equalities over nodes. Given a proof with such sequents, we show that all axiomatic sequents can be assigned interpolants---which are themselves sequents---and that such interpolants can be `propagated' through the proof yielding 
an interpolant of the conclusion. Explicit definitions can then be readily extracted from these interpolants. We note that our method is constructive in the sense that interpolants are computed relative to a \emph{given proof} of a general concept inclusion implied by a $\riq$-ontology. Although such proofs are in principle computable, we left the specification of an explicit proof-search algorithm that builds such proofs to future work, noting that such algorithms can be written by adapting known techniques; e.g.~\cite{HorSat04}. 

Finally, we remark that although our work shares similarities with that of~\cite{TenFraSey13}, our method goes beyond this work as we establish the CBP for the highly expressive DL $\riq$, and due to the modularity of our sequent systems, our method is applicable to restrictions of $\riq$. By \emph{modularity} we mean that the deletion of inference rules or modification of side conditions on rules allows for sequent systems to be provided for fragments of $\riq$. Our work also intersects that of \cite{CatConMarVen06}, which establishes the CBP for $\alc$ extended with PUR Horn conditions, but differs both in terms of methodology and that our work covers extensions of $\alcq$. 

\textbf{Outline of Paper}: In Section~\ref{sec:log-prelims}, we define the 
logic $\riq$, define the CBP and related notions, and explicate certain grammar theoretic concepts used in formulating inference rules. In Section~\ref{sec:sequent-systems}, we present our sequent systems and establish that each system enjoys fundamental 
properties. Section~\ref{sec:interpolants} develops and explains our new sequent-based method that computes explicit definitions of implicitly definable concepts and establishes the CBP, using $\riq$ as case study. To the best of our knowledge, this is the first proof that $\riq$ has the CBP. In Section~\ref{sec:conclusion}, we conclude and discuss future work. We note that all proofs have been deferred to the appendix.






%% file: body-1.tex
In the first part of this section, we introduce the language and semantics for the description logic $\riq$, which subsumes various DLs
~\cite{CalDeG03}. Subsequently, we discuss and define a 
notion of interpolation and concept-based Beth definability, which will be of pivotal interest in this paper. In the last part of this section, we introduce special types of semi-Thue systems~\cite{Pos47}, referred to as \emph{$\roles$-systems}, which are essential in the formulation of our sequent calculi. 

\subsection{Language and Semantics: $\riq$}

The description logic $\riq$ is defined relative to a \emph{vocabulary} $\vocab = (\rnames, \cnames)$, which is a pair containing pairwise disjoint, countable sets. The set $\rnames$ contains \emph{role names} used to denote binary relations and the set $\cnames$ contains \emph{concept names} used to denote classes of entities. We use the (potentially annotated) symbols $\role$, $\roleb$, $\ldots$ to denote role names, and $A$, $B$, $\ldots$ to denote concept names. We define a \emph{role} to be a role name or an \emph{inverse role} $\invn{\role}$ such that $\role \in \rnames$. We define the \emph{inverse} of a role to be $\inv{\role} = \invn{\role}$ and $\inv{\invn{\role}} = \role$ given that $\role \in \rnames$. We let $\roles := \rnames \cup \{\inv{\role} \ | \ \role \in \rnames\}$ denote the set of roles.

 A \emph{complex role inclusion axiom (RIA)} is an expression $\role_{1} \circ \cdots \circ \role_{n}\imp \roleb$ such that $\role_{1}, \ldots, \role_{n}$ and $\roleb$ are roles, and $\circ$ denotes the usual composition operation over binary relations; we assume $n$-ary compositions $\role_{1} \circ \cdots \circ \role_{n}$ associate to the left. We define an \emph{RBox} $\rbox$ to be a finite collection of RIAs. As identified by Horrocks and Sattler~\cite{HorSat04}, to ensure the decidability of reasoning with $\riq$, only \emph{regular RBoxes} may be used in ontologies (defined below).\footnote{Note, our interpolation results go through for general RBoxes, i.e. this restriction is not needed for the work in Sections~\ref{sec:sequent-systems} and \ref{sec:interpolants}.} 
  Let $\prec$ be a strict partial order on the set $\rnames$ of role names; we define an RIA $w \imp \role$ to be \emph{$\prec$-regular} \iffi $\role$ is a role name, and either (1) $w = \role\role$, (2) $w = \role^-$, (3) $w = \roleb_1 \circ \cdots \circ \roleb_n$ and $\roleb_i \prec \role$ for all $1 \leq i \leq n$, (4) $w = \role \circ \roleb_1 \circ \cdots \circ \roleb_n$ and $\roleb_i \prec \role$ for all $1 \leq i \leq n$, or (5) $w = \roleb_1 \circ \cdots \circ \roleb_n \circ \role$ and $\roleb_i \prec \role$ for all $1 \leq i \leq n$. An RBox $\rbox$ is defined to be \emph{regular} \iffi a strict partial order $\prec$ over $\rnames$ exists such that every RIA in $\rbox$ is $\prec$-regular. 
 
We recursively define a role name $\role$ to be \emph{simple} (with respect to an RBox $\rbox$) \iffi either (1) no RIA of the form $w \imp \role$ occurs in $\rbox$, or (2) for each $\roleb \imp \role \in \rbox$, $\roleb$ is a simple role name or its inverse is. Also, an inverse role $\invn{\role}$ is defined to be \emph{simple} if $\role$ is simple.
 

 We define \emph{complex concepts} to be formulae in negation normal form generated by the following grammar in BNF:
$$
C ::= A \ | \ \neg A \ | \ (C \odot C) \ | \ (Q . C) \ | \ ({\leqslant} n \roleb.C) \ | \ ({\geqslant} n \roleb.C)
$$
where $A \in \cnames$, $\odot \in \{\dis, \con\}$, $Q \in \{\exists \role, \forall \role \ | \ \role \in \roles \}$, $\roleb$ is a simple role, and $n \in \mathbb{N}$. We use the symbols $C$, $D$, $\ldots$ (potentially annotated) to denote complex concepts. We define $\top = A \dis \neg A$ and $\bot = A \con \neg A$ for a fixed $A \in \cnames$, and define a \emph{literal} $L$ to be either a concept name or its negation, i.e. $L \in \{A, \neg A \ | \ A \in \cnames \}$. 
 For a concept name $A$, we define $\negnnf{A} := \neg A$ and $\negnnf{\neg A} := A$, and 
 we lift the definition of negation to complex concepts in the usual way, noting that 
 $\negnnf{({\leqslant} n \role.C)} := ({\geqslant} (n{+}1) \role. C)$, and 
\[
  \negnnf{({\geqslant} n \role. C)} :=
  \begin{cases}
     \bot & \text{if $n=0$,} \\
     {\leqslant} (n{-}1) \role. C & \text{otherwise.}
  \end{cases}
\]
We recursively define the \emph{weight} of a concept $C$ as follows: 
 (1) $\wght{L} = 1$ with $L \in \{A, \neg A \ | \ A \in \cnames \}$, (2) $\wght{C \odot D} = \wght{C} + \wght{D} + 1$ with $\odot \in \{\con, \dis\}$, (3) $\wght{Q.C} = \wght{C} + 1$ with $Q \in \{\exists \role, \forall \role \ | \ \role \in \roles \}$, (4) $\wght{{\leqslant} n \roleb.C} = \wght{C} + n + 1$, and (5) $\wght{{\geqslant} n \roleb.C} = \wght{C} + n$.

 A \emph{general concept inclusion axiom (GCI)} is a formula of the form $C \sqsubseteq D$ such that $C$ and $D$ are complex concepts. A \emph{TBox} $\tbox$ is a finite set of GCIs and we make the simplifying assumption that every GCI in a TBox $\tbox$ is of the form $\top \sqsubseteq C$.  We define a \emph{$\riq$-ontology} $\ont$ (which we refer to as an \emph{ontology} for short) to be the union of an RBox $\rbox$ and TBox $\tbox$, that is, $\ont = \rbox \cup \tbox$. For a set $X$ of concepts, GCIs, or RIAs, we let $\cpt{X}$ denote the set of all concept names occurring in $X$, and we let $\sig{X}$ denote the set of all concept names and roles occurring in $X$. Symbols from a vocabulary $\vocab$ are interpreted accordingly:
 


\begin{definition}[Interpretation]\label{def:DL-interpretation} An \emph{interpretation} $\inter = (\dom,\map{\cdot})$ is a pair consisting of a non-empty set $\dom$ called the \emph{domain} and a map $\map{\cdot}$ such that
\begin{description}


\item[$\bullet$] if $A \in \cnames$, then $\map{A} \subseteq \dom$ with $\map{\neg A} = \dom \setminus \map{A}$;

\item[$\bullet$] if $\role \in \rnames$, then $\map{\role} \subseteq \dom \times \dom$.

\end{description}
 We define $\map{(\invn{\role})} = \{(b,a) \ | \ (a,b) \in \map{\role}\}$ and interpret compositions over roles in the usual way. 
We lift interpretations to complex concepts accordingly: 
\begin{description}

\item[$\bullet$] $\map{(C \sqcup D)} = \map{C} \cup \map{D}$;

\item[$\bullet$] $\map{(C \sqcap D)} = \map{C} \cap \map{D}$;

\item[$\bullet$] $\map{\exists \role. C} = \{a \in \dom \ | \ \exists b \in \dom,  (a,b) \in \map{\role} \ \& \ b \in \map{C}\}$;

\item[$\bullet$] $\map{\forall \role. C} = \{a \in \dom \ | \ \forall b \in \dom, (a,b) \in \map{\role} \Rightarrow b \in \map{C}\}$;

\item[$\bullet$] $\map{\ltn C} {:=} \{a {\in} \dom \ : \ \card{\{b \ : \ (a,b) \in \map{\roleb} \ \& \ b \in \map{C} \}} \leq n \}$;

\item[$\bullet$] $\map{\gtn C} {:=} \{a {\in} \dom \ : \ \card{\{b \ : \ (a,b) \in \map{\roleb} \ \& \ b \in \map{C} \}} \geq n \}$.

\end{description}
An interpretation \emph{satisfies} $C \imp D$ or $\role_{1} \circ \cdots \circ \role_{n} \imp \roleb$, written $\inter \vDash C \sqsubseteq D$ and $\inter \vDash \role_{1} \circ \cdots \circ \role_{n}\imp \roleb$ \iffi $\map{C} \subseteq \map{D}$ and $\map{(\role_{1} \circ \cdots \circ \role_{n})} \subseteq \map{\roleb}$, respectively. An interpretation $\inter$ is defined to be a \emph{model} of an ontology $\ont$, written $\inter \vDash \ont$, \iffi it satisfies all GCIs and RIAs in $\ont$. We write $\ont \vDash C \imp D$ \iffi for every interpretation $\inter$, if $\inter \vDash \ont$, then $\inter \vDash C \imp D$, and we write $\ont \vDash C \equiv D$ when $\ont \vDash C \sqsubseteq D$ and $\ont \vDash D \sqsubseteq C$.
\end{definition}

\subsection{Definability and Interpolation}\label{subsec:def-interp}

The notion of 
Beth definability, first defined within the context of first-order logic~\cite{Bet56}, takes on a number of distinct formulations within the context of DLs. In~\cite{BaaNut03,CatConMarVen06}, Beth definability is reinterpreted as the notion of \emph{definitorial completeness}, which has also been named \emph{concept-based Beth definability (CBP)}~\cite{TenFraSey13}. Intuitively, a DL $\deslog$ has the CBP when the implicit definability of a concept $C$ under an $\deslog$-ontology $\ont$ using a signature $\sigtr = \sigtrii \cup \rnames$ with $\sigtrii \subseteq \cpt{C,\ont}$ implies its explicit definability using symbols from $\sigtr$. This is distinct from the \emph{projective Beth definability property (PBDP)}, which is defined in the same way but relative to a signature $\sigtr \subseteq \sig{C,\ont}$, or the weaker \emph{Beth definability property (BDP)} where the signature $\sigtr$ is the set of all symbols distinct from the concept defined~\cite{ArtJunMazOzaWol23}. In this paper, we focus on the CBP, and leave the investigation of sequent-based methodologies for establishing other definability properties to future work. Let us now formally define the CBP.

Let $\deslog$ be a DL, $C$ be a complex concept in $\deslog$, $\ont$ an $\deslog$-ontology, and $\sigtr \subseteq \cpt{C,\ont}$.\footnote{In this paper, we take a DL $\deslog$ to be $\riq$ or a fragment thereof.} We define $C$ to be \emph{implicitly concept-definable} from $\sigtr$ under $\ont$ \iffi for any two models $\inter$ and $\interii$ of $\ont$ such that $\dom = \domii$ and for each $P \in \sigtr \cup \rnames$, $P^{\inter} = P^{\interii}$, it follows that $C^{\inter} = C^{\interii}$. We remark that this notion can be reformulated as a standard reasoning problem, that is, $C$ is implicitly concept-definable from $\sigtr$ under $\ont$ \iffi 
\begin{equation}\label{eq:implicit-def}
\ont \cup \ont_{\sigtr} \vDash C \imp C_{\sigtr}
\end{equation}
where $\ont_{\sigtr}$ and $C_{\sigtr}$ are obtained from $\ont$ and $C$, respectively, by uniformly replacing every concept name $A \not\in \sigtr$ by a fresh concept name. We define $C$ to be \emph{explicitly concept-definable} from $\sigtr$ under $\ont$ \iffi there exists a complex concept $D$ (called an \emph{explicit concept-definition}) such that $\ont \vDash C \equiv D$ and $\cpt{D} \subseteq \sigtr$.

\begin{definition}[Concept-Based Beth Definability] Let $\deslog$ be a DL, $C$ be a complex concept in $\deslog$, $\ont$ be an $\deslog$-ontology, and $\sigtr \subseteq \cpt{C,\ont}$. We say that $\deslog$ has the \emph{concept-name Beth definability property (CBP)} \iffi if $C$ is implicitly concept-definable from $\sigtr$ under $\ont$, then $C$ is explicitly concept-definable from $\sigtr$ under $\ont$. 
\end{definition}

It is typical to establish definability properties by means of an interpolation theroem (cf.~\cite{TenFraSey13,Cra57,JunMazWol22}). We therefore define a suitable notion of interpolation that implies the CBP, which we call \emph{concept interpolation}.

\begin{definition}[Concept Interpolation Property]\label{def:concept-interpolant} Let $\deslog$ be a DL, $\ont_{1}$ and $\ont_{2}$ be $\mathcal{L}$-ontologies with $\ont = \ont_{1} \cup \ont_{2}$, and $C$ and $D$ be $\mathcal{L}$-concepts. We define an $\mathcal{L}$-concept $I$ to be a \emph{concept interpolant} for $C \imp D$ under $\ont$ \iffi (1) $\cpt{I} \subseteq \cpt{\ont_{1}, C} \cap \cpt{\ont_{2}, D}$, (2) $\ont \vDash C \imp I$, and (3) $\ont \vDash I \imp D$. A DL $\mathcal{L}$ enjoys the \emph{concept interpolation property} if for all $\mathcal{L}$-ontologies $\ont_{1}$, $\ont_{2}$ with $\ont = \ont_{1} \cup \ont_{2}$ and $\mathcal{L}$-concepts $C$, $D$ such that $\ont \vDash C \sqsubseteq D$ there exists a concept interpolant for $C \sqsubseteq D$ under $\ont$.
\end{definition}

\begin{lemma}\label{lem:interpolation-implies-CBP} If a DL $\deslog$ enjoys the concept interpolation property, then it enjoys the CBP.
\end{lemma}

\subsection{$\roles$-Systems}

 We let $\roles$ serve as our \emph{alphabet} with each role serving as a \emph{character}. The set $\roles^{*}$ of \emph{strings over $\roles$} is defined to be the smallest set satisfying the following conditions: (i) $\roles \cup \{\varepsilon\} \subseteq \roles^{*}$ with $\varepsilon$ the \emph{empty string}, and (ii) $\text{If } \stra \in \roles^{*} \text{ and } \role \in \albet \text{, then } \stra \concat \role \in \roles^{*}$, where $\stra \concat \role$ represents the \emph{concatenation} of $\stra$ and $\role$. We use $\stra, \strb, \ldots$ (potentially annotated) to denote strings from $\albetstr$, and we have $\stra \cate \empstr = \empstr \cate \stra = \stra$, for the empty string $\empstr$. The inverse operation on strings is defined as: (1) $\inv{\varepsilon} := \varepsilon$, and (2) $\text{If } \stra = \role_{1} \cdots \role_{n} \text{, then } \inv{\stra} := \inv{\role_{n}} \cdots \inv{\role_{1}}$.

We now define \emph{$\roles$-systems}, which are special types of \emph{Semi-Thue systems}~\cite{Pos47}, relative to ontologies. These will permit us to derive strings of roles from a given role and encode the information present in a given ontology. 

\begin{definition}[$\roles$-system]\label{def:grammar} Let $\ont$ be an ontology. We define the \emph{$\roles$-system} $\sto{\ont}$ to be the smallest set of \emph{production rules} of the form $\role \pto \stra$, where $\role \in \roles$ and $\stra \in \albetstr$, such that if $\role_{1} \circ \cdots \circ \role_{n} \imp \roleb \in \ont$, then
$$
(\roleb \pto \role_{1} \cate \cdots \cate \role_{n}), (\inv{\roleb} \pto \inv{\role_{n}} \cate \cdots \cate \inv{\role_{1}}) \in \sto{\ont}.
$$
\end{definition}


\begin{definition}[Derivation, Language]\label{def:semi-thue-deriv-lang}
 Let $\ont$ be an ontology and $\sto{\ont}$ be its $\roles$-system. We write $\stra \pto_{\sto{\ont}} \strb$ and say that the string $\strb$ may be derived from the string $\stra$ in \emph{one-step} \iffi there are strings $\stra', \strb' \in \albetstr$ and $\role \pto \strc \in \sto{\ont}$ such that $\stra = \stra' \cate \role \cate \strb'$ and $\strb = \stra' \cate \strc \cate \strb'$. We define the \emph{derivation relation} $\dr$ to be the reflexive and transitive closure of $\pto_{\sto{\ont}}$. For $\stra, \strb \in \albetstr$, we call $\stra \dr \strb$ a \emph{derivation of $\strb$ from $\stra$}, and define the \emph{length} of a derivation to be the minimal number of one-step derivations required to derive $\strb$ from $\stra$ in $\sto{\ont}$. Last, we define the \emph{language} $\glang(\role) := \{\stra \ | \ \role \dr \stra \}$, where $\role \in \albet$.
\end{definition}



%% file: body-2.tex
We let $\lab = \{\la, \lb, \lc, \ldots\}$ be a countably infinite set of labels, define a \emph{role atom} to be an expression of the form $\role (\la,\lb)$ with $\role \in \roles$ and $\la, \lb \in \lab$, define an \emph{equality atom} and \emph{inequality atom} to be an expression of the form $\la \eq \lb$ and $\la \noteq \lb$ with $\la, \lb \in \lab$, respectively, and define a \emph{labeled concept} to be an expression $\la : C$ with $\la \in \lab$ and $C$ a complex concept. We refer to role, equality, and inequality atoms as \emph{structural atoms} more generally. For a (multi)set $X$ and $Y$ of structural atoms and/or labeled concepts, we let $X,Y$ represent their union and let $\lab(X)$ be the set of labels occurring therein. We say that a set $\rel$ of structural atoms \emph{forms a tree} \iffi the graph $T(\rel) = (V,E)$ is a directed tree with $V = \lab(\rel)$, and $(x,y) \in E$ \iffi $\role(x,y) \in \rel$. A \emph{sequent} is defined to be an expression of the form $\seq := \rel \sar \cxti$ such that (1) $\rel$ is a set of structural atoms that forms a tree, (2) $\cxti$ is a multiset of labeled concepts, (3) if $\rel \neq \emptyset$, then $\lab(\cxti) \subseteq \lab(\rel)$, and (4) if $\rel = \emptyset$, then $|\lab(\cxti)| = 1$. In a sequent $\rel \sar \cxti$, we refer to $\rel$ as the \emph{antecedent}, $\cxti$ as the \emph{consequent}, 
 and we define $\cxti {\restriction} x := \{C \ | \ x : C \in \cxti\}$. 

Recall that every GCI in an ontology $\ont$ is assumed to be of the form $\top \imp C$. For an ontology $\ont = \rbox \cup \tbox$ and label $\la \in \lab$, we let $\la : \gcilist{\ont} = \la : \negnnf{C}_{1}, \ldots, \la : \negnnf{C}_{n}$ such that $\tbox = \{\top \imp C_{1}, \ldots, \top \imp C_{n} \}$. For labels $\la_{1}, \ldots, \la_{n} \in \lab$, we define $\neqset{\la_{1}, \ldots, \la_{n}} = \{\la_{i} \noteq \la_{j} \ | \ 1 \leq i < j \leq n\}$. We let $x \approx y \in \{x \eq y, y \eq x\}$ and write $\eqpath{x}{y}$ \iffi there exist $z_{1}, \ldots, z_{n} \in \lab(\rel)$ such that $z_{1} \approx z_{2}, \ldots, z_{n-1} \approx z_{n}$ with $x = z_{1}$ and $y = z_{n}$. We make use of equivalence classes of labels in the formulation of certain inference rules below and define $\eqclass{x}{\rel} := \{y \ | \ \eqpath{x}{y}\}$ for a sequent $\rel \sar \cxti$.

\begin{figure*}[t] 

\begin{center}
\begin{tabular}{c c c c}
\AxiomC{}
\RightLabel{$\idc$}
\UnaryInfC{$\rel \sar \la : A, \la : \neg A, \cxti$}
\DisplayProof

&

\AxiomC{}
\RightLabel{$\ideq^{\dag_{1}}$}
\UnaryInfC{$\rel, x \noteq y \sar \cxti$}
\DisplayProof

&

\AxiomC{$\rel \sar \la : L, \lb : L, \cxti$}
\RightLabel{$\subeq^{\dag_{1}}$}
\UnaryInfC{$\rel \sar \la : L, \cxti$}
\DisplayProof

&

\AxiomC{$\rel \sar \la : C, \la : D, \cxti$}
\RightLabel{$\disr$}
\UnaryInfC{$\rel \sar \la : C \dis D, \cxti$}
\DisplayProof
\end{tabular}
\end{center}

\begin{center}
\begin{tabular}{c c c}
\AxiomC{$\rel \sar \la : C, \cxti$}
\AxiomC{$\rel \sar \la : D, \cxti$}
\RightLabel{$\conr$}
\BinaryInfC{$\rel \sar \la : C \con D, \cxti$}
\DisplayProof

&

\AxiomC{$\rel \sar \la : \some C, \lb : C, \cxti$}
\RightLabel{$\existsr^{\dag_{2}}$}
\UnaryInfC{$\rel \sar \la : \some C, \cxti$}
\DisplayProof

&

\AxiomC{$\rel, \role(\la,\lb) \sar \lb : C, \lb : \gcilist{\ont}, \cxti$}
\RightLabel{$\allr^{\dag_{3}}$}
\UnaryInfC{$\rel \sar \la : \all C, \cxti$}
\DisplayProof
\end{tabular}
\end{center}

\begin{center}
\begin{tabular}{c c}
\AxiomC{$\rel' \sar \lb_{0} : \negnnf{C}, \lb_{0} : \gcilist{\ont}, \ldots, \lb_{n} : \negnnf{C}, \lb_{n} : \gcilist{\ont}, \cxti$}
\RightLabel{$\ltnr^{\dag_{4}}$}
\UnaryInfC{$\rel \sar \la : {\leqslant} n \role . C, \cxti$}
\noLine
\UnaryInfC{$\rel' = \rel, \neqset{y_{0}, \ldots, y_{n}}, \role(\la,\lb_{0}), \ldots, \role(\la,\lb_{n})$}
\DisplayProof

&

\AxiomC{$\rel \sar \lb_{i} : C, \la : {\geqslant} n \role . C, \cxti \  | \ 1 \leq i \leq n$}
\noLine
\UnaryInfC{$\rel, y_{i} \eq y_{j} \sar \la : {\geqslant} n \role . C, \cxti \ | \ 1 \leq i < j \leq n$}
\RightLabel{$\gtnr^{\dag_{5}}$}
\UnaryInfC{$\rel \sar \la : {\geqslant} n \role . C, \cxti$}
\DisplayProof
\end{tabular}
\end{center}

\medskip

\noindent
\textbf{Side Conditions:}

\medskip

\begin{minipage}{.45\textwidth}
\begin{description}

\item[$\dag_{1} =$] `$\eqpath{x}{y}$.'  

\item[$\dag_{2} =$] `$\prgr{\rel} \vDash \eqclass{x}{\rel} \prpath{L} \eqclass{y}{\rel}$ with $L = \glang(\role)$.'

\item[$\dag_{3} =$] `$y$ is fresh.'

\end{description}
\end{minipage}
\begin{minipage}{.45\textwidth}
\begin{description}

\item[$\dag_{4} =$] `For each $0 \leq i \leq n$, $y_{i}$ is fresh.'

\item[$\dag_{5} =$] `For each $1 \leq i \leq n$, $\prgr{\rel} \vDash \eqclass{x}{\rel} \prpath{L} \eqclass{y_{i}}{\rel}$ with $L = \glang(\role)$.' 

\end{description}
\end{minipage}


\caption{The calculus $\calc$ for the $\riq$-ontology $\ont$. The rules with side conditions $\dag_{1}$ -- $\dag_{5}$ are applicable only if that side condition holds.}
\label{fig:calculus}
\end{figure*}

A uniform presentation of our sequent systems in presented in \fig~\ref{fig:calculus}. We note that each sequent calculus $\calc$ takes a $\riq$-ontology $\ont$ as an input parameter, which determines the functionality of certain inference rules depending on the contents of $\ont$. The calculus $\calc$ contains the \emph{initial rules} $\id$ and $\ideq$, which generate axioms that are used to begin a proof, the \emph{logical rules} $\disr$, $\conr$, $\existsr$, $\allr$, $\ltnr$, and $\gtnr$, which introduce complex concepts, and the \emph{substitution rule} $\subeq$. We note that $A \in \cnames$ in the $\id$ rule and $L$ is a literal in the $\subeq$ rule. The $\ideq$ and $\subeq$ rules are subject to a side condition, namely, each rule is applicable only if $\eqpath{x}{y}$. The $\allr$ and $\ltnr$ rules are subject to side conditions as well: the label $y$ and the labels $y_{0}, \ldots, y_{n}$ must be \emph{fresh} in $\allr$ and $\ltnr$, respectively, meaning such labels may not occur in the conclusion of a rule application. Last, we note that the $\existsr$ and $\gtnr$ rules are special types of logical rules, referred to as \emph{propagation rules}; cf.~\cite{CasCerGasHer97,Fit72}. These rules operate by viewing sequents as types of automata, referred to as \emph{propagation graphs}, which bottom-up propagate formulae along special paths, referred to as \emph{propagation paths} (see Example~\ref{ex:prop-rule} below). 

\begin{definition}[Propagation Graph]\label{def:propagation-graph} We define the \emph{propagation graph} $\prgr{\rel} = (V,E)$ of a sequent $\rel \sar \cxti$ such that $\eqclass{x}{\rel} \in V$ \iffi $x \in \lab(\rel)$, and $(\eqclass{x}{\rel}, \role, \eqclass{y}{\rel}), (\eqclass{y}{\rel}, \inv{\role}, \eqclass{x}{\rel}) \in E$ \iffi there exist $z \in \eqclass{x}{\rel}$ and $w \in \eqclass{y}{\rel}$ such that $\role(z, w) \in \rel$. If we write $\eqclass{x}{\rel} \in \prgr{\rel}$, then we mean $\eqclass{x}{\rel} \in \prgrdom$, and if we write $(\eqclass{x}{\rel},\role,\eqclass{y}{\rel}) \in \prgr{\rel}$, we mean $(\eqclass{x}{\rel},\role,\eqclass{y}{\rel}) \in \prgredges$.
\end{definition}

We note that our propagation graphs are generalizations of those employed in sequent systems for modal and non-classical logics~\cite{CiaLyoRamTiu20,GorPosTiu11,Lyo21b}. In particular, due to the inclusion of equality atoms, we must define propagation graphs over equivalence classes of labels, rather than over labels themselves. This lets us define novel and correct propagation rules in the presence of (in)equalities and counting quantifiers. 

\begin{definition}[Propagation Path]\label{def:propagation-path} 
Given a propagation graph $\prgr{\rel} = (V,E)$, $\eqclass{x}{\rel}, \eqclass{y}{\rel} \in V$, and $\role \in \roles$, we write $\prgr{\rel} \vDash \eqclass{x}{\rel} \prpath{\role} \eqclass{y}{\rel}$ \iffi $(\eqclass{x}{\rel},\role,\eqclass{y}{\rel}) \in E$. Given a string $\role\stra \in \albetstr$ where $\role \in \roles$, we define $\prgr{\rel} \vDash \eqclass{x}{\rel} \prpath{\role\stra} \eqclass{y}{\rel}$ as `$\exists_{\eqclass{z}{\rel} \in V} \ \prgr{\rel}  \vDash \eqclass{x}{\rel} \prpath{\role} \eqclass{z}{\rel}$ and $\prgr{\rel} \vDash \eqclass{z}{\rel} \prpath{\stra} \eqclass{y}{\rel}$', and we take $\prgr{\rel} \vDash \eqclass{x}{\rel} \prpath{\empstr} \eqclass{y}{\rel}$ to mean that $\eqclass{x}{\rel} = \eqclass{y}{\rel}$. Additionally, when $\prgr{\rel}$ is clear from the context we may simply write $\eqclass{x}{\rel} \prpath{\stra} \eqclass{y}{\rel}$ to express $\prgr{\rel} \vDash \eqclass{x}{\rel} \prpath{\stra} \eqclass{y}{\rel}$. Finally, given a language $\glang(\role)$ of some $\roles$-system $\sto{\ont}$ and $\role \in \albet$, we use $\eqclass{x}{\rel} \prpath{L} \eqclass{y}{\rel}$ with $L = \glang(\role)$ \iffi  there is a string $\stra \in \glang(\role)$ such that $\eqclass{x}{\rel} \prpath{\stra} \eqclass{y}{\rel}$.
\end{definition}

To provide intuition concerning the functionality of propagation rules, we illustrate a (bottom-up) application of $\gtnr$. 

\begin{example}\label{ex:prop-rule} Let us consider the sequent $\Gamma \sar x : {\geqslant} 2 \role . C$ with $\Gamma = r(x,y), r(x,z), r(x,w), z \eq w$. A pictorial representation of the propagation graph $\prgr{\Gamma}$ is shown below.
\begin{center}
\begin{tikzpicture}		
\node[] (x) [] {$\{y\}$};
\node[] (y0) [right=of x,xshift=1cm] {$\{x\}$};
\node[] (y1) [right=of y0,xshift=1cm] {$\{z,w\}$};

\draw [->, dotted](y0) to [bend right=25] node[midway,above] {$r$} (x);
\draw [->,dotted](x) to [bend right=25] node[midway,below] {$\inv{r}$} (y0);
\draw [->,dotted](y0) to [bend left=25] node[midway,above] {$r$} (y1);
\draw [->,dotted](y1) to [bend left=25] node[midway,below] {$\inv{r}$} (y0);
\end{tikzpicture}
\end{center}
One can see that there are two labels $y$ and $z$ such that $\eqclass{x}{\rel} \prpath{r} \eqclass{y}{\rel}$ and $\eqclass{x}{\rel} \prpath{r} \eqclass{z}{\rel}$. Note that $r \in \glang(\role)$ by definition. Therefore, we may (bottom-up) apply the $\gtnr$ rule to obtain the three premises $\Gamma \sar x : {\geqslant} 2 \role . C, y : C$, $\Gamma \sar x : {\geqslant} 2 \role . C, z : C$, and $\Gamma, y \eq z \sar x : {\geqslant} 2 \role . C$.
\end{example}


We define a \emph{proof} in $\calc$ inductively: (1) each instance of an initial rule $(r)$, as shown below left, is a proof with conclusion $\seq$, and (2) if $n$ proofs exist with the respective conclusions $\seq_{1}$, $\ldots$, $\seq_{n}$, then applying an $n$-ary rule $(r')$, as shown below right, yields a new proof with conclusion $\seq$.
\begin{center}
\begin{tabular}{c c}
\AxiomC{}
\RightLabel{$(r)$}
\UnaryInfC{$\seq$}
\DisplayProof

&

\AxiomC{$\seq_{1} \ \cdots \ \seq_{n}$}
\RightLabel{$(r')$}
\UnaryInfC{$\seq$}
\DisplayProof
\end{tabular}
\end{center}
We use $\prf$ (potentially annotated) to denote proofs, and we say a sequent $\seq$ is \emph{provable} with $\prf$ in $\calc$, written $\calc, \prf \Vdash \seq$ \iffi $\seq$ is the conclusion of $\prf$. We write $\calc \Vdash \seq$ to indicate that $\seq$ is provable with some $\prf$ in $\calc$. 
 Observe that each proof is a tree of sequents with the conclusion as the root. We define the \emph{height} of a proof to be the number of sequents along a maximal branch from the conclusion to an initial rule of the proof. The \emph{size} of a proof $\prf$ is defined to be the sum of the weights of the sequents it contains; in other words, $\size{\prf} := \sum_{S \in \prf} \wght{S}$, where the \emph{weight} of a sequent $\seq = \rel \sar \cxti$ is defined to be $\wght{\seq} := |\rel| + \sum_{\la : C \in \cxti} \wght{C}$. Ignoring labeled concepts of the form $x : \gcilist{\ont}$, we refer to the formulae that are explicitly mentioned in the premises of a rule as \emph{active}, and those explicitly mentioned in the conclusion as \emph{principal}. For example, $r(x,y)$ and $y : C$ are active in $\allr$ while $x : \all C$ is principal.

\begin{figure*}[t] 

\begin{center}
\begin{tabular}{c c c c}
\AxiomC{ }
\RightLabel{$\topr$}
\UnaryInfC{$\rel \sar \cxti, x : \top$}
\DisplayProof

&

\AxiomC{$\rel \sar \cxti$}
\RightLabel{$\relab{x}{y}^{\dag_{1}}$}
\UnaryInfC{$\rel(x/y) \sar \cxti(x/y)$}
\DisplayProof

&

\AxiomC{$\rel \sar \cxti$}
\RightLabel{$\wkeq^{\dag_{2}}$}
\UnaryInfC{$\rel, x \eq y \sar \cxti$}
\DisplayProof

&

\AxiomC{$\rel \sar \cxti$}
\RightLabel{$\wkneq^{\dag_{2}}$}
\UnaryInfC{$\rel, x \noteq y \sar \cxti$}
\DisplayProof
\end{tabular}
\end{center}

\begin{center}
\begin{tabular}{c c c}
\AxiomC{$\rel \sar \cxti$}
\RightLabel{$\wk^{\dag_{3}}$}
\UnaryInfC{$\rel \sar \cxti, x : C$}
\DisplayProof

&

\AxiomC{$\rel \sar \cxti, x : C, x : C$}
\RightLabel{$\ctr$}
\UnaryInfC{$\rel \sar \cxti, x : C$}
\DisplayProof

&

\AxiomC{$\rel, x \noteq y, y \noteq x  \sar \cxti$}
\RightLabel{$\symneq$}
\UnaryInfC{$\rel, x \noteq y \sar \cxti$}
\DisplayProof
\end{tabular}
\end{center}

\caption{(Hp-)admissible rules in $\calc$. The 
 side conditions are: $\dag_{1}$ = `$x$ is fresh,' $\dag_{2}$ = `$x,y \in \lab(\rel, \cxti)$,' and $\dag_{3}$ = `$x \in \lab(\rel, \cxti)$.'}
\label{fig:admiss-rules}
\end{figure*}

We now define a semantics for our sequents, which is used to establish our sequent systems sound and complete.

\begin{definition}[Sequent Semantics]\label{def:sequent-semantics} Let $\inter = (\dom,\map{\cdot})$ be an interpretation, $\seq = \rel \sar \cxti$ a sequent, $\lmap : \lab(\rel,\cxti) \to \dom$ a \emph{label assignment}, and $\ont$ an ontology.
\begin{description}

\item[$\bullet$] $\inter, \lmap \amodels \rel$ \iffi for each $\role(\la,\lb), \la \eq \lb, \la \noteq \lb \in \rel$, we have $(\lmap(\la),\lmap(\lb)) \in \map{\role}$, $\lmap(\la) = \lmap(\lb)$, and $\lmap(\la) \neq \lmap(\lb)$;

\item[$\bullet$] $\inter, \lmap \emodels \cxti$ \iffi for some $\la : C \in \cxti$, $\lmap(\la) \in \map{C}$.

\end{description}
 A sequent $\seq = \rel \sar \cxti$ is \emph{satisfied} in $\inter$ with $\lmap$ relative to $\ont$, written $\inter, \lmap \omodels \seq$, \iffi if $\inter \vDash \ont$ and $\inter, \lmap \amodels \rel$, then $\inter, \lmap \emodels \cxti$. A sequent $\seq = \rel \sar \cxti$ is \emph{true} in $\inter$ relative to $\ont$, written $\inter \omodels \seq$, \iffi $\inter, \lmap \omodels \seq$ for all label assignments $\lmap$. A sequent $\seq = \rel \sar \cxti$ is \emph{valid} relative to $\ont$, written $\omodels \seq$, \iffi $\inter \omodels \seq$ for all interpretations $\inter$, and we say that $\seq$ is \emph{invalid} relative to $\ont$ otherwise, writing $\not\omodels \seq$.
\end{definition}

\begin{lemma}\label{lem:ppath-implies-R}
Let $\inter = (\dom,\map{\cdot})$ be an interpretation, $\ont$ be a $\riq$ ontology, $\lambda$ be a label assignment, and $\rel$ be a set of structural atoms. If $\inter \vDash \ont$, $\inter, \lmap \amodels \rel$, and $\prgr{\rel} \vDash \eqclass{x}{\rel} \prpath{L} \eqclass{y}{\rel}$ with $L = \glang(\role)$, then $(\lambda(x), \lambda(y)) \in \map{\role}$.
\end{lemma}


\begin{theorem}[Soundness]\label{thm:soundness}
If $\calc \Vdash \rel \vdash \cxti$, then $\omodels \rel \sar \cxti$. 
\end{theorem}



We now confirm that $\calc$ enjoys desirable proof-theoretic properties, viz. certain rules are height-preserving admissible or invertible. A rule is \emph{(height-preserving) admissible, i.e. (hp-)admissible}, if the premises of the rule have proofs (of heights $h_{1}, \ldots, h_{n}$), then the conclusion of the rule has a proof (of height $h \leq \max\{h_{1}, \ldots, h_{n}\}$). If we let $(r^{-1})$ be the inverse of the rule $(r)$ whose premise is the conclusion of $(r)$ and conclusion is the premises of $(r)$, then we say that $(r)$ is \emph{(height-preserving) invertible, i.e. (hp-)invertible} \iffi $(r^{-1})$ is (hp-)admissible. For a sequent $\seq = \rel \sar \cxti$, we let $\seq(x/y) = \rel(x/y) \sar \cxti(x/y)$ denote the sequent obtained by substituting each occurrence of the label $y$ in $\seq$ by $x$; for example, if $\seq = r(x,y), x \neq y \sar y : A$, then $\seq(z/y) = r(x,z), x \neq z \sar z : A$. Important (hp-)admissible rules are displayed in \fig~\ref{fig:admiss-rules}.


\begin{lemma}\label{prop:hp-admiss-wk}
The $\topr$ rule is provable in $\calc$, and the $\relab{x}{y}$, $\wkeq$, $\wkneq$, $\wk$, $\ctr$, and $\symneq$ rules are hp-admissible. 
\end{lemma}

\begin{lemma}\label{prop:hp-invert}
All non-initial rules in $\calc$ are hp-invertible.
\end{lemma}

The completeness of $\calc$ (stated below) is shown by taking a sequent of the form $\emptyset \sar \la : \gcilist{\ont}, \la : C$ as input and showing that if the sequent is not provable, then $\calc$ can be used to construct a counter-model thereof, witnessing the invalidity of the sequent relative to $\ont$. 

\begin{theorem}[Completeness]\label{thm:completeness}
If $\omodels \emptyset \sar \la : C$, then $\calc \Vdash \emptyset \sar \la : \gcilist{\ont}, \la : C$. 
\end{theorem}

The following corollary is a consequence of \thm~\ref{thm:soundness} and \thm~\ref{thm:completeness}. We write $\calc \Vdash C \imp D$ as shorthand for $\calc \Vdash \emptyset \sar x : \gcilist{\ont}, x : \negnnf{C} \dis D$.

\begin{corollary}\label{cor:subsumption-sequent-equiv}
$\ont \vDash C \imp D$ \iffi $\calc \Vdash C \imp D$.
\end{corollary}


Last, we emphasize the modularity of our sequent systems and approach. By omitting inference rules for certain connectives and/or only accepting certain ontologies as the input parameter $\ont$, sequent calculi can be obtained for DLs serving as fragments of $\riq$; cf.~\cite{CalDeG03}. For example, sequent calculi for $\alc$ ontologies are easily obtained by omitting the $\ideq$, $\subeq$, $\gtnr$, and $\ltnr$ rules. 
 The constructive method presented next applies to fragments of $\riq$ by leveraging this feature, 
 thus demonstrating its generality. 


%% file: body-3.tex
We now describe our methodology for computing concept interpolants, and by extension, explicit concept-definitions of implicitly defined concepts (by \lem~\ref{lem:interpolation-implies-CBP}). The central idea is to generalize the notion of a concept interpolant from GCIs to sequents. Then, given a proof of a sequent $\seq$, we assign concept interpolants to all initial sequents of the proof, and show how a concept interpolant can be defined for the conclusion of a rule application from those of its premises, culminating in a concept interpolant for $\seq$. As sequents are more general than GCIs, this approach will establish, in a constructive manner, that $\riq$ (and its various sublogics) enjoy the concept interpolation property and the CBP.

\begin{definition}[Interpolant]\label{def:sequent-interpolant} We define an \emph{interpolant} to be a set $\itp := \{\rel_{i} \sar \cxti_{i} \ | \ 1 \leq i \leq n \}$ such that $\rel_{i}$ is a set of (in)equalities of the form $x \eq y$ and $x \noteq y$ with $x, y \in \lab$ distinct labels, and $\cxti_{i}$ is a set of labeled concepts. Given an interpolant $\itp$ of the above form, we define its \emph{orthogonal} $\orth{\itp}$ as follows: $\rel \sar \cxti \in \orth{\itp}$ \iffi for each $1 \leq i \leq |\rel, \cxti|$, one and only one of the following holds: (1) $x \eq y \in \rel$ with $x \noteq y \in \rel_{i}$, (2) $x \noteq y \in \rel$ with $x \eq y \in \rel_{i}$, or (3) $x : \negnnf{C} \in \cxti$ with $x : C \in \cxti_{i}$. We use $\itp$ and annotated versions 
for interpolants.
\end{definition}

\begin{example} Let $\itp = \{(x \eq y \sar x : A), (z \noteq u \sar z : \neg B) \}$. Then, the orthogonal $\orth{\itp}$ is the set containing $(x \noteq y, z \eq u \sar \ )$, $(x \noteq y \sar z \! : \! B )$, $(z \eq u \sar x \! : \! \neg A)$, and $(\ \sar x \! : \! \neg A, z \! : \! B)$, that is, each member of $\orth{\itp}$ is formed by including a negated element from each member of $\itp$.
\end{example}

In order to fully specify our interpolant construction algorithm, we need to define two special interpolants, named $\all \itp$ and $\leqnr . \itp$, which appear in quantifier and qualified number restriction rules. We let $\clist$ denote a set of complex concepts, define $x : \clist := \{x : C \ | \ C \in \clist \}$, and define $\bigsqcap \clist$, $\bigsqcup \clist$, and $\negnnf{\clist}$ to be the conjunction, disjunction, and negation of all complex concepts in $\clist$, respectively. 

\begin{definition}\label{def:special-interpolants} Let $\itp = \{ \rel \sar \cxti_{i}, y : \clist_{i} \ | \ 1 \leq i \leq m\}$ such that $y \not\in \lab(\rel)$ and $\cxti_{i} {\restriction} y = \emptyset$, then we define:
$$
\all \itp := \{ \rel \sar \cxti_{i}, \la : \all \bigsqcup \clist_{i} \ | \ 1 \leq i \leq m\}.
$$  
Let $\itp = \{ \rel, \rel' \sar \cxti_{i}, y_{0} : \clist_{0,i}, \ldots, y_{n} : \clist_{n,i} \ | \ 1 \leq i \leq m\}$ such that $\lab(\rel) \cap \{y_{0}, \ldots, y_{n}\} = \emptyset$, $\rel' \subseteq \neqset{y_{0}, \ldots, y_{n}}$, $\cxti_{i} {\restriction} y_{j} = \emptyset$ for $0 \leq j \leq n$, and $\clist_{i} = \clist_{0,i}, \ldots, \clist_{n,i}$. Then,
$$
\leqnr . \itp := \{ (\rel \sar \cxti_{i}, \la : \leqnr .  \negnnf{\bigsqcup \clist_{i}}) \ | \ 1 \leq i \leq m\}.
$$
\end{definition}

An \emph{interpolation sequent} is defined to be an expression of the form $\rel; \Phi \splt \Psi \sar \cxti \splt \cxtii \sep \itp$ such that $\rel$ is a set of role and equality atoms, $\Phi, \Psi$ is a set of inequality atoms, $\cxti, \cxtii$ is a multiset of labeled concepts, $\itp$ is an interpolant, and $a, b \in \{1,2\}$ with $a \neq b$. For an interpolation sequent of the aforementioned form, we refer to $\rel, \Phi \sar \cxti$ as the \emph{left partition} and $\rel, \Psi \sar \cxtii$ as the \emph{right partition}. Recall that for a concept interpolant $I$ of a GCI $C \imp D$ under $\ont$, the ontology $\ont$ is the union of two ontologies $\ont_{1}$ and $\ont_{2}$ such that $\cpt{I} \subseteq \cpt{\ont_{1}, C} \cap \cpt{\ont_{2}, D}$ (see \dfn~\ref{def:concept-interpolant}). The use of $a, b \in \{1,2\}$ in an interpolation sequent is to keep track of which partition is associated with which ontology, 
 e.g. in $\rel; \Phi \splti \Psi \sar \cxti \splti \cxtii \sep \itp$ the left (right) partition is associated with $\ont_{1}$ ($\ont_{2}$, respectively).

\begin{figure*}[t] 

\begin{center}
\begin{tabular}{c c}
\AxiomC{}
\RightLabel{$\idcIi$}
\UnaryInfC{$\rel; \Phi \splt \Psi \sar \cxti, \la : A \splt \la : \neg A, \cxtii \sep \{(\ \sar \la : \neg A)\}$}
\DisplayProof

&

\AxiomC{}
\RightLabel{$\idcIii$}
\UnaryInfC{$\rel; \Phi \splt \Psi  \sar \cxti \splt \la : A, \la : \neg A, \cxtii \sep \{(\ \sar \la : \top)\}$}
\DisplayProof
\end{tabular}
\end{center}

\begin{center}
\AxiomC{}
\RightLabel{$\ideqI$}
\UnaryInfC{$\rel; \Phi \splt \Psi, x \noteq y  \sar \cxti \splt \cxtii \sep \{ (x \noteq y \sar \ ) \}$}
\DisplayProof
\end{center}

\begin{center}
\begin{tabular}{c c}
\AxiomC{$\rel; \Phi \splt \Psi \sar \cxti \splt \cxtii \sep \itp$}
\RightLabel{$\orthru$}
\UnaryInfC{$\rel; \Psi \spltrev \Phi \sar \cxtii \spltrev \cxti \sep \orth{\itp}$}
\DisplayProof

&

\AxiomC{$\rel, \role(\la,\lb); \Phi \splt \Psi \sar \cxti \splt \lb : C, \lb : \gcilist{\ont}, \cxtii \sep \itp$}
\RightLabel{$\allrI$}
\UnaryInfC{$\rel; \Phi \splt \Psi \sar \cxti \splt \la : \all C, \cxtii \sep \all \itp$}
\DisplayProof
\end{tabular}
\end{center}

\begin{center}
\begin{tabular}{c}
\AxiomC{$\rel, \role(\la,\lb_{0}), \ldots, \role(\la,\lb_{n}); \Phi \splt \Psi, \neqset{y_{0}, \ldots, y_{n}} \sar \cxti \splt \lb_{0} : \negnnf{C}, \lb_{0} : \gcilist{\ont}, \ldots, \lb_{n} : \negnnf{C}, \lb_{n} : \gcilist{\ont}, \cxtii \sep \itp$}
\RightLabel{$\ltnrI$}
\UnaryInfC{$\rel; \Phi \splt \Psi \sar \cxti \splt \la : {\leqslant} n \role . C, \cxtii \sep \leqnr . \itp$}
\DisplayProof
\end{tabular}
\end{center}

\caption{Rules in $\icalc$. The $\ideqI$, $\allrI$, and $\ltnrI$ rules satisfy the same side conditions as $\ideq$, $\allr$, and $\ltnr$, respectively.}
\label{fig:interpolation-rules-1}
\end{figure*}

\begin{definition}[Interpolant Preserving Rules] Let $\ru$ be a rule in the set $\{\subeq, \disr, \conr, \existsr, \gtnr\}$ of the form shown below and assume that the active equalities and/or labeled concepts occur in $\rel_{i}$ and/or $\cxtii_{i}$, respectively, with the principal formula in $\cxtii$.
\begin{center}
\AxiomC{$\rel_{i}, \Phi, \Psi \sar \cxti, \cxtii_{i} \ | \ 1 \leq i \leq n$}
\RightLabel{$\ru$}
\UnaryInfC{$\rel, \Phi, \Psi \sar \cxti, \cxtii$}
\DisplayProof
\end{center}
We define its corresponding \emph{interpolant rule} as follows:
\begin{center}
\begin{center}
\AxiomC{$\rel_{i}; \Phi \splt \Psi \sar \cxti \splt \cxtii_{i} \sep \itp_{i} \ | \ 1 \leq i \leq n $}
\RightLabel{$\ruI$}
\UnaryInfC{$\rel; \Phi \splt \Psi \sar \cxti \splt \cxtii \sep \itp_{1} \cup \cdots \cup \itp_{n}$}
\DisplayProof
\end{center}
\end{center}
We refer to a rule $\subeqI$, $\disrI$, $\conrI$, $\existsrI$, or $\gtnrI$ as an \emph{interpolant preserving rule}, or \emph{IP-rule}. We stipulate that $\existsrI$ and $\gtnrI$ are subject to the same side conditions as $\existsr$ and $\gtnr$, respectively, w.r.t. the propagation graph $\prgr{\rel}$.
\end{definition}

For each sequent calculus $\calc$, we define its corresponding \emph{interpolation calculus} accordingly:
$$
\icalc := \{\idcIi, \idcIii, \orthru\} \cup \{\ruI \ | \ \ru \in \calc \setminus \id \}
$$
Observe that interpolation calculi contain IP-rules as well as rules from \fig~\ref{fig:interpolation-rules-1}. In an interpolation calculus $\icalc$, the $\idcIi$, $\idcIii$, and $\ideqI$ rules are the \emph{initial rules}, $\orthru$ is the \emph{orthogonal rule}, $\subeqI$ is the \emph{substitution rule}, and all remaining rules are \emph{logical rules}. The orthogonal rule cuts the number of rules needed in $\icalc$ roughly in half as it essentially `swaps' the left and right partition permitting rules to be defined that only operate within the right partition; cf.~\cite{LyoTiuGorClo20}. A \emph{proof}, its \emph{height}, and the provability relation $\Vdash$ are defined in $\icalc$ in the same manner as for $\calc$. 

We now put forth a sequence of lemmas culminating in the main interpolation theorem (\thm~\ref{thm:interpolation}), which implies that $\riq$ has the CBP (\cor~\ref{cor:RIQ-cbp}). We remark that Lemmas~\ref{lem:main-interpolation-lemma-1} and \ref{lem:interpolation-lemma} describe proof transformation algorithms between $\calc$ and $\icalc$. In particular, Lemma~\ref{lem:main-interpolation-lemma-1} states that each proof in $\calc$ of a sequent $\rel, \Phi, \Psi \sar \cxti, \cxtii$ in a special form can be transformed into a proof  in $\icalc$ of a specific interpolation sequent $\rel; \Phi \splt \Psi \sar \cxti \splt \cxtii \sep \itp$. Then, via Lemma~\ref{lem:interpolation-lemma}, this proof can be transformed into two proofs in $\calc$ witnessing that the interpolant $\itp$ is `implied by' the left partition $\rel, \Phi \sar \cxti$ and `implies' the right partition $\rel, \Psi \sar \cxtii$. Both Lemmas~\ref{lem:main-interpolation-lemma-1} and \ref{lem:interpolation-lemma} are shown by induction on the height of the given proof. Last, when we use the notation $\rel, \Phi_{a}, \Phi_{b} \sar \gcilist{\ont_{a}}, \cxti_{a}, \cxti_{b}, \gcilist{\ont_{b}}$ or the notation $\rel; \Phi_{a} \splt \Phi_{b} \sar \gcilist{\ont_{a}}, \cxti_{a} \splt \cxti_{b}, \gcilist{\ont_{b}} \sep \itp$, we assume that $\gcilist{\ont_{c}} := x_{1} : \gcilist{\ont_{c}}, \ldots, x_{n} : \gcilist{\ont_{c}}$ such that $\lab(\rel, \Phi_{c}, \cxti_{c}) = \{x_{1}, \ldots, x_{n}\}$ and $c \in \{a,b\}$. The use of $\gcilist{\ont_{c}}$ ensures each partition satisfies its respective ontology. 

\begin{lemma}\label{lem:main-interpolation-lemma-1}
Let $\ont = \ont_{1} \cup \ont_{2}$ be an ontology 
and suppose that $\rel, \Phi, \Psi \sar \gcilist{\ont_{a}}, \cxti, \cxtii, \gcilist{\ont_{b}}$ has a proof $\prf$ in $\calc$ with $\Phi \cap \Psi = \emptyset$. Then, $\prf$ can be transformed into a proof in $\icalc$ of $\rel; \Phi \splt \Psi \sar \gcilist{\ont_{a}}, \cxti \splt \cxtii, \gcilist{\ont_{b}} \sep \itp$ such that:
\begin{itemize}

\item[(1)] If $x \eq y$ occurs in $\itp$, then $x \noteq y \in \Phi$;

\item[(2)] If $x \noteq y$ occurs in $\itp$, then $x \noteq y \in \Psi$;

\item[(3)] $\lab(\itp) \subseteq \lab(\rel, \Phi, \Psi, \gcilist{\ont_{a}}, \cxti, \cxtii, \gcilist{\ont_{b}})$; 

\item[(4)] $\cpt{\itp} \subseteq \cpt{\ont_{a}, \cxti} \cap \cpt{\cxtii, \ont_{b}}$.
 
\end{itemize}
\end{lemma}

The following lemma states that a double orthogonal transformation always `preserves' some of the sequents from the original interpolant. As shown in the appendix, the lemma is helpful in proving \lem~\ref{lem:interpolation-lemma}.

\begin{lemma}\label{lem:orth-lemma}
If $(\cxtii \sar \cxtiii) \in \orth{\orth{\itp}}$, then there exists a $(\rel \sar \cxti) \in \itp$ such that $\rel \subseteq \cxtii$ and $\cxti \subseteq \cxtiii$. 
\end{lemma}

 

\begin{lemma}\label{lem:interpolation-lemma}
If $\icalc \Vdash \rel; \Phi \splt \Psi \sar \cxti \splt \cxtii \sep \itp$, then 
\begin{itemize}

\item[(1)] For each $(\rel' \sar \cxtiii_{i}) \in \itp$, $\calc \Vdash \rel, \rel', \Phi \sar \cxti, \cxtiii_{i}$; 

\item[(2)] For each $(\rel' \sar \cxtiv_{i}) \in \orth{\itp}$, $\calc \Vdash \rel, \rel', \Psi \sar \cxtiv_{i}, \cxtii$. 

\end{itemize}
\end{lemma}

Next, we prove that an interpolant containing at most a single label, i.e. an interpolant of the form
$$
\itp := \{( \ \sar x : C_{i,1}, \ldots, x : C_{i,k_{i}}) \ | \ 1 \leq i \leq n\}
$$
can be transformed into a single labeled concept within the context of a proof. Toward this end, we define $x : \bigsqcap \bigsqcup \itp :=  x: \bigsqcap_{1 \leq i \leq n} \bigsqcup_{1 \leq j \leq k_{i}} C_{i,j}$, where $\itp$ is as above. The following two lemmas are straightforward and follow by applying the $\disr$ and $\conr$ rules in $\calc$ a sufficient number of times.

\begin{lemma}
\label{lem:sub-interpolation-lemma-3}
If $\rel \sar \cxti, \cxtii$ is provable in $\calc$ for all $(\ \vdash \cxtii) \in \itp$ and $\lab(\itp) = \{\la\}$, then $\calc \Vdash \rel \sar \cxti, \la : \bigsqcap \bigsqcup \itp$.
\end{lemma}

\begin{lemma}
\label{lem:sub-interpolation-lemma-4}
If $\rel \sar \cxti, \cxtii$ is provable in $\calc$ for all $(\ \sar \cxtii) \in \orth{\itp}$ and $\lab(\itp) = \{\la\}$, then $\calc \Vdash \rel \sar \cxti, \la : \negnnf{\bigsqcap \bigsqcup \itp}$. 
\end{lemma}

Our main theorem below is a consequence of Lemmas~\ref{prop:hp-invert}--\ref{lem:sub-interpolation-lemma-4}. Given a proof of $\sar x : \gcilist{\ont}, x : \negnnf{C} \dis D$, we obtain proofs of $\sar x : \gcilist{\ont_{1}}, x : \negnnf{C}, x : I$ and $\sar x : \gcilist{\ont_{2}}, x : D, x : \negnnf{I}$ in $\calc$ by Lemmas~\ref{prop:hp-invert}, \ref{lem:main-interpolation-lemma-1}, and \ref{lem:interpolation-lemma}--\ref{lem:sub-interpolation-lemma-4} with $I = \bigsqcap \bigsqcup \itp$. The concept interpolant $I$ is computed in $\exptime$ due to the potential use of the $\orthru$ rule, which may exponentially increase the size of interpolants.

\begin{theorem}\label{thm:interpolation}
Let $\ont = \ont_{1} \cup \ont_{2}$ be a $\riq$ ontology. If $\ont \vDash C \imp D$, i.e. $\calc, \prf \Vdash C \imp D$, then a concept interpolant $I$ can be computed in $\exptime$ relative to $\size{\prf}$ such that $\calc \Vdash C \imp I$ and $\calc \Vdash I \imp D$, i.e. $\ont \vDash C \imp I$ and $\ont \vDash I \imp D$.
\end{theorem}

Let $C$ be a complex concept, $\ont$ be a $\riq$ ontology, $\sigtr \subseteq \cpt{C,\ont}$, and suppose $C$ is implicitly concept-definable from $\sigtr$ under $\ont$. If we want to find the explicit concept-definition of $C$ from $\sigtr$ under $\ont$, we utilize the sequent calculus $\mathsf{S}(\ont')$ with $\ont' = \ont \cup \ont_{\sigtr}$. Since $C$ is implicitly concept-definable, we know by (1) in Section~\ref{subsec:def-interp} and Corollary~\ref{cor:subsumption-sequent-equiv} that $\mathsf{S}(\ont') \Vdash C \imp C_{\sigtr}$. By applying Theorem~\ref{thm:interpolation}, we obtain a concept interpolant $I$ for $C \imp C_{\sigtr}$ under $\ont$, which serves as an explicit concept-definition by \lem~\ref{lem:interpolation-implies-CBP}. Therefore, we have a constructive proof of the following corollary.


\begin{corollary}\label{cor:RIQ-cbp}
$\riq$ has the concept interpolation property and the CBP.
\end{corollary}


%% file: conclusion.tex

We have provided novel sequent calculi for $\riq$ ontologies, showing them sound and complete, as well as showing that each calculus enjoys useful hp-admissibility and hp-invertibility properties. Our sequent systems are modular as the omission of certain rules or constructs from ontologies yields sequent systems for restrictions of $\riq$. Moreover, we presented a sequent-based methodology for computing concept interpolants and explicit definitions of implicitly definable concepts using $\riq$ as a case study, thus satisfying a demand for developing constructive interpolation and definability methods. To the best of our knowledge, we have provided the first proof of the CBP for $\riq$.

There are various avenues of future research. First, it would be interesting to know the size and complexity of computing a concept interpolant $I$ relative to $C \imp D$ rather than from a proof $\prf$ witnessing $\ont \vDash C \imp D$. This can be achieved by supplying a proof-search algorithm that generates a proof of $C \imp D$, whose relative complexity and size can then be determined. Second, we aim to generalize our methodology to decide and compute the existence of \emph{Craig interpolants} for $\riq$ and related DLs, which is a non-trivial problem (see~\cite{TenFraSey13}). Last, we could  generalize our method to consider constructs beyond those in $\riq$, e.g. negations over roles, intersections of roles, nominals, or the @ operator; it is known that for some of these extensions, e.g. nominals, even concept interpolation fails~\cite{ArtJunMazOzaWol23}, requiring an increase in complexity to decide the existence of interpolants.

%% file: app-0.tex
\begin{customlem}{\ref{lem:interpolation-implies-CBP}} If a DL $\deslog$ enjoys the concept interpolation property, then it enjoys the CBP.
\end{customlem}

\begin{proof} Let $\ont$ be an $\mathcal{L}$-ontology, $C$ be an $\mathcal{L}$-concept, and $\sigtr \subseteq \cpt{C, \ont}$. Suppose $\mathcal{L}$ enjoys the concept interpolation property and let $C$ be implicitly concept-definable from $\sigtr$ under $\ont$, that is, $\ont \cup \ont_{\sigtr} \vDash C \imp C_{\sigtr}$ (see (\ref{eq:implicit-def}) on p.~\pageref{eq:implicit-def}). Then, a concept interpolant $I$ exists such that (i) $\cpt{I} \subseteq \cpt{\ont, C} \cap \cpt{\ont_{\sigtr}, C_{\sigtr}}$, (ii) $\ont \cup \ont_{\sigtr} \vDash C \imp I$, and (iii) $\ont \cup \ont_{\sigtr} \vDash I \imp D$. By (i) and the definitions of $\ont_{\sigtr}$ and $C_{\sigtr}$, we have that $\cpt{I} \subseteq \sigtr$. We now establish that $\ont \vDash C \equiv I$. We argue that $\ont \vDash C \imp I$, and note that the argument showing $\ont \vDash I \imp C$ is similar.

Suppose $\inter \vDash \ont$. We aim to show that $\inter \vDash C \imp I$. First, recall that each concept name $B \in \cpt{C, \ont}$ such that $B \not\in \sigtr$ is replaced by a fresh concept name in $C_{\sigtr}$ and $\ont_{\sigtr}$, which we denote by $B'$. Now, let us define $\interii = (\domii, \cdot^{\interii})$ such that $\Delta^{\interii} := \dom$, for each $B \in \cpt{C, \ont}$ such that $B \not\in \sigtr$, $(B')^{\interii} := \map{B}$, and for all other symbols $P$, let $P^{\interii} := \map{P}$. Observe that $\interii \vDash \ont \cup \ont_{\sigtr}$, meaning $\interii \vDash C \imp I$ by (ii) above. Furthermore, observe that for $P \in \cpt{C, \ont} \cup \rnames$, $P^{\interii} = \map{P}$, meaning $\inter \vDash C \imp I$ as $C$ and $I$ are only composed of symbols from $\cpt{C, \ont} \cup \rnames$. It can be argued in a similar fashion that $\ont \vDash I \imp C$, meaning, $\ont \vDash C \equiv I$. 

Therefore, $I$ serves as an explicit concept-definition of $C$, showing that $C$ is explicitly concept-definable from $\sigtr$ under $\ont$, that is, $\dl$ has the CBP.
\end{proof}

%% file: app-1.tex
\begin{customlem}{\ref{lem:ppath-implies-R}}
Let $\inter = (\dom,\map{\cdot})$ be an interpretation, $\ont$ be a $\riq$ ontology, $\lambda$ be a label assignment, and $\rel$ be a set of structural atoms. If $\inter \vDash \ont$, $\inter, \lmap \amodels \rel$, and $\prgr{\rel} \vDash \eqclass{x}{\rel} \prpath{L} \eqclass{y}{\rel}$ with $L = \glang(\role)$, then $(\lambda(x), \lambda(y)) \in \map{\role}$.
\end{customlem}

\begin{proof} Suppose $\inter \vDash \ont$, $\inter, \lmap \amodels \rel$, and $\eqclass{x}{\rel} \prpath{L} \eqclass{y}{\rel}$ with $L = \glang(\role)$. We prove the claim by induction on the length of the derivation of the string $\stra \in L$ such that $\eqclass{x}{\rel} \prpath{\stra} \eqclass{y}{\rel}$.

\textit{Base Case.} For the base case, we suppose that the derivation of $\stra$ is of length $0$. By \dfn~\ref{def:semi-thue-deriv-lang}, we know that the only derivation in $\gram(\ont)$ from $\role$ of length $0$ is the derivation of the form $\role$, that is, the derivation consisting solely of $\role$. Therefore, $\eqclass{x}{\rel} \prpath{\role} \eqclass{y}{\rel}$, which implies that either $\role(z,w) \in \rel$ or $\inv{\role}(w, z) \in \rel$ for $z \in \eqclass{x}{\rel}$ and $w \in \eqclass{y}{\rel}$ by \dfn~\ref{def:propagation-graph}. In either case, since $\inter, \lmap \amodels \rel$, we have that $(\lambda(x), \lambda(y)) \in \map{\role}$.

\textit{Inductive Step.} Let the derivation of $\stra$ be of length $n+1$. By \dfn~\ref{def:semi-thue-deriv-lang}, there is a derivation $\role \dr \strb \roleb \strc$ of length $n$ and production rule $\roleb \pto \roleb_{1} \cdots \roleb_{m} \in \gram(\ont)$ such that $\stra = \strb \roleb_{1} \cdots \roleb_{m} \strc$. As $\stra$ is a string encoding a propagation path in $\prgr{\rel}$, it follows that 
 $\eqclass{\lc_{1}}{\rel} \prpath{\roleb_{1}} \eqclass{\lc_{2}}{\rel} \cdots \eqclass{\lc_{m}}{\rel} \prpath{\roleb_{m}} \eqclass{\lc_{m+1}}{\rel}$ occurs in $\prgr{\rel}$. By \dfn~\ref{def:propagation-graph}, $\hat{\roleb}_{1}(z_{1}',z_{2}'), \ldots, \hat{\roleb}_{m}(z_{m}',z_{m+1}') \in \rel$ with $z_{i}' \in \eqclass{\lc_{i}}{\rel}$ and
$$
\hat{\roleb}_{i+1}(z_{i}',z_{i+1}') \in \{\roleb_{i+1}(z_{i}',z_{i+1}'), \inv{\roleb_{i+1}}(z_{i+1}',z_{i}')\}
$$
 for $1 \leq i \leq m$. This implies that $(\lambda(z_{1}),\lambda(z_{m+1})) \in \map{(s_{1} \circ \cdots \circ s_{m})}$ because $\inter, \lmap \amodels \rel$. Furthermore, since $s \pto s_{1} \cdots s_{m} \in \gram(\ont)$, either $s_{1} \circ \cdots \circ s_{m} \imp s \in \ont$ or $\inv{s_{m}} \circ \cdots \circ \inv{s_{1}} \imp \inv{s} \in \ont$, by \dfn~\ref{def:grammar}. Regardless of the case, $(\lambda(z_{1}),\lambda(z_{m+1})) \in \map{s}$. We will use this fact to complete the proof of the inductive step below.
 
Let us consider the set $\rel' = \rel, \roleb(z_{1}',z_{m+1}')$ of relational atoms. We know that $\inter, \lmap \amodels \rel'$ because $\inter', \lmap \amodels \rel$ by assumption and $(\lambda(z_{0}),\lambda(z_{m+1})) \in \map{s}$ by what was argued above. Moreover, we have that $\prgr{\rel'} \vDash \eqclass{x}{\rel'} \prpath{\strb} \eqclass{z_{1}}{\rel'} \prpath{\roleb} \eqclass{z_{m+1}}{\rel'} \prpath{\strc} \eqclass{y}{\rel'}$. Observe that $\strb \roleb \strc$ has a derivation of length $n$ by what was said above. Therefore, we may invoke the induction hypothesis, from which it follows that $(\lambda(x), \lambda(y)) \in \map{\role}$.
\end{proof}

\begin{customthm}{\ref{thm:soundness}}[Soundness]
If $\calc \Vdash \rel \vdash \cxti$, then $\omodels \rel \sar \cxti$. 
\end{customthm}

\begin{proof} By induction on the height of the proof of $\rel \vdash \cxti$.\\

\noindent
\textit{Base Case.} If the height of the proof is $1$, then our proof is an instance of $\id$ or $\ideq$, as shown below.
\begin{center}
\begin{tabular}{c c}
\AxiomC{}
\RightLabel{$\idc$}
\UnaryInfC{$\rel \sar \la : A, \la : \neg A, \cxti$}
\DisplayProof

&

\AxiomC{}
\RightLabel{$\ideq$}
\UnaryInfC{$\rel, x \noteq y \sar \cxti$}
\DisplayProof
\end{tabular}
\end{center}
We argue the $\id$ case first, and assume for a contradiction that $\rel \sar \la : A, \la : \neg A, \cxti$ is invalid relative to $\ont$. From this, it follows that there exists an interpretation $\mathcal{I}$ and a label assignment $\lambda$ such that $\mathcal{I}, \lambda \amodels \rel$ and $\mathcal{I}, \lambda \not\emodels \la : A, \la : \neg A, \cxti$. Hence, $\lambda(x) \notin A^\mathcal{I}$ and $\lambda(x) \in A^\mathcal{I}$, which is a contradiction. For the $\ideq$ case, we assume for a contradiction that $\rel, x \noteq y \sar \cxti$ is invalid relative to $\ont$. Then, there exists an interpretation $\mathcal{I}$ and label assignment $\lmap$ such that $\mathcal{I}, \lambda \amodels \rel, x \noteq y$ and $\mathcal{I}, \lambda \not\emodels \cxti$. By the side condition $\eqpath{x}{y}$, we have that $\lmap(x) = \lmap(y)$. However, we also have that $\lmap(x) \neq \lmap(y)$, which is a contradiction.\\

\noindent
\textit{Inductive Step}. Assume soundness holds for proofs of height $n$. We now show that it holds for proofs of height $n + 1$. We prove each case by contraposition and argue that if the conclusion of the last inference of the proof is invalid relative to $\ont$, then at least one premise of the inference must be invalid relative to $\ont$.\\ 

\noindent
The $\subeq$ rule:
\begin{center}
\AxiomC{$\rel \sar \la : L, \lb : L, \cxti$}
\RightLabel{$\subeq$}
\UnaryInfC{$\rel \sar \la : L, \cxti$}
\DisplayProof
\end{center}
Suppose there exists an interpretation $\mathcal{I}$ and label assignment $\lmap$ such that $\mathcal{I}, \lambda \amodels \rel$ and $\mathcal{I}, \lambda \not\emodels \la : L, \cxti$. By the side condition $\eqpath{x}{y}$ imposed on $\subeq$, we know that $\lmap(x) = \lmap(y)$. Since we have that $\lmap(\la) \not\in \map{L}$, it follows that $\lmap(\lb) \not\in \map{L}$, which shows the premise invalid relative to $\ont$.\\

\noindent
The $\disr$ rule:
\begin{center}
\begin{tabular}{c c}
\AxiomC{$\rel \sar \la : C, \la : D, \cxti$}
\RightLabel{$\disr$}
\UnaryInfC{$\rel \sar \la : C \dis D, \cxti$}
\DisplayProof
\end{tabular}
\end{center}
 Assume $\rel \sar \la : C \dis D, \cxti$ is invalid relative to $\ont$. By \dfn~\ref{def:sequent-semantics}, it follows that there exists an interpretation $\mathcal{I}$ and a label assignment $\lambda$ such that $\mathcal{I}, \lambda \amodels \rel$ and $\mathcal{I}, \lambda \not\emodels \la : C \dis D, \cxti$. Hence, $\lambda(x) \notin \map{C}$ and $\lambda(x) \notin \map{D}$. Moreover, it holds that $\mathcal{I}, \lambda \not\emodels \cxti$. Since $\mathcal{I}, \lambda \vDash \rel$ and $\mathcal{I}, \lambda \not\emodels \la : C, \la : D, \cxti$, the premise is invalid relative to $\ont$ as well.\\

\noindent
The $\conr$ rule:
\begin{center}
\begin{tabular}{c c}
\AxiomC{$\rel \sar \la : C, \cxti$}
\AxiomC{$\rel \sar \la : D, \cxti$}
\RightLabel{$\conr$}
\BinaryInfC{$\rel \sar \la : C \con D, \cxti$}
\DisplayProof
\end{tabular}
\end{center}
Assume $\rel \sar \la : C \con D, \cxti$ is invalid relative to $\ont$. By \dfn~\ref{def:sequent-semantics}, it follows that there exists an interpretation $\mathcal{I}$ and a label assignment $\lambda$ such that $\mathcal{I}, \lambda \amodels \rel$ and $\mathcal{I}, \lambda \not\emodels \la : C \con D, \cxti$. Either $ \lambda(x) \notin \map{C}$ or $\lambda(x) \notin \map{D}$. Also, $\mathcal{I}, \lambda \not\emodels \cxti$. Therefore, either $\mathcal{I}, \lambda \not\emodels \la : C, \cxti$ or $\mathcal{I}, \lambda \not\emodels \la : D, \cxti$, meaning, at least one of the premises must be invalid relative to $\ont$ as well.\\

\noindent
The $\existsr$ rule:
\begin{center}
\AxiomC{$\rel \sar \la : \some C, \lb : C, \cxti$}
\RightLabel{$\existsr$}
\UnaryInfC{$\rel \sar \la : \some C, \cxti$}
\DisplayProof
\end{center}
Suppose there exists an interpretation $\mathcal{I}$ and label assignment $\lmap$ such that $\mathcal{I}, \lambda \amodels \rel$ and $\mathcal{I}, \lambda \not\emodels \la : \some C, \cxti$. By the side condition imposed on $\existsr$, we know that $\eqclass{x}{\rel} \prpath{L} \eqclass{y}{\rel}$ with $L = \glang(\role)$. By \lem~\ref{lem:ppath-implies-R}, we know that $(\lambda(x), \lambda(y)) \in \map{\role}$. As $\lmap(x) \notin \map{\some C}$, we know that $\lmap(y) \notin \map{C}$, showing that the premise is invalid relative to $\ont$.\\

\noindent
The $\allr$ rule:
\begin{center}
\AxiomC{$\rel, \role(\la,\lb) \sar \lb : C, \lb : \gcilist{\ont}, \cxti$}
\RightLabel{$\allr$}
\UnaryInfC{$\rel \sar \la : \all C, \cxti$}
\DisplayProof
\end{center}
Suppose, $\rel \sar \la : \all C, \cxti$ is invalid relative to $\ont$. From this, it follows that there exists an interpretation $\mathcal{I}$ and a label assignment $\lambda$ such that $\inter \vDash \ont$, $\mathcal{I}, \lambda \amodels \rel$ and $\mathcal{I}, \lambda \not\emodels \la : \all C, \cxti$.
Thus, there is at least one domain element $a \in \dom$ such that $(\lambda(x),a) \in \map{\role}$ and $a \notin \map{C}$. 
We now define a new label assignment $\lambda'$ such that $\lambda'(z) = \lambda(z)$ if $z \neq y$ and $\lambda'(y) = a$.
Hence, $\mathcal{I}, \lambda' \amodels  \rel, r(\la, \lb)$ and $\lambda'(y) \notin \map{C}$. Moreover, let $y_{i} : \gcilist{\ont} = y : \negnnf{D}_{1}, \ldots y : \negnnf{D}_{k}$. Since $\top \imp D_{j} \in \ont$ for $1 \leq j \leq k$, we have that $\inter, \lambda' \not\emodels y : \gcilist{\ont}$ as $\inter \vDash \ont$. Therefore, the premise is invalid relative to $\ont$. \\

\noindent
The $\ltnr$ rule, where $\rel' = \rel, \neqset{y_{0}, \ldots, y_{n}}, \role(\la,\lb_{0}), \ldots, \role(\la,\lb_{n})$:
\begin{center}
\AxiomC{$\rel' \sar \lb_{0} : \negnnf{C}, \lb_{0} : \gcilist{\ont}, \ldots, \lb_{n} : \negnnf{C}, \lb_{n} : \gcilist{\ont}, \cxti$}
\RightLabel{$\ltnr$}
\UnaryInfC{$\rel \sar \la : {\leqslant} n \role . C, \cxti$}
\DisplayProof
\end{center}
Assume, $\rel \sar \la : {\leqslant} n \role . C, \cxti$ is invalid relative to $\ont$. From this, it follows that there exists an interpretation $\mathcal{I}$ and a label assignment $\lambda$ such that $\inter \vDash \ont$, $\mathcal{I}, \lambda \amodels \rel$ and $\mathcal{I}, \lambda \not\emodels \la : {\leqslant} n \role . C, \cxti$. It follows that $\lmap(\la) \in \map{({\geqslant} (n{+}1) \role. C)}$. Thus, there are at least $n+1$ many distinct elements $a_{0}, \ldots, a_{n} \in \dom$ such that $(\lambda(x),a_{i}) \in \map{\role}$ and $a_{i} \in \map{C}$. We now define a new label assignment $\lambda'$ such that $\lambda'(z) = \lambda(z)$ if $z \neq y_i$ with $0 \leq i \leq n$ and $\lambda'(y_i) = a_i$. Hence, $\mathcal{I}, \lambda' \amodels  \rel, \neqset{y_{0}, \ldots, y_{n}},  \role(\la,\lb_{0}), \ldots, \role(\la,\lb_{n})$ and $\mathcal{I}, \lambda' \not\emodels \lb_{0} : \negnnf{C}, \ldots, \lb_{n} : \negnnf{C}, \cxti$ as $\lambda'(y_i) \in \map{C}$ for each $0 \leq i \leq n$. Moreover, let $y_{i} : \gcilist{\ont} = y_{i} : \negnnf{D}_{1}, \ldots y_{i} : \negnnf{D}_{k}$ for $0 \leq i \leq n$. Since $\top \imp D_{j} \in \ont$ for $1 \leq j \leq k$, we have that $\inter, \lambda \not\emodels y_{0} : \gcilist{\ont}, \ldots, y_{n} : \gcilist{\ont}$ as $\inter \vDash \ont$. Therefore, the premise is invalid relative to $\ont$.\\

\noindent
The $\gtnr$ rule:
\begin{center}
\AxiomC{$\rel \sar \lb_{i} : C, \la : {\geqslant} n \role . C, \cxti \ | \ 1 \leq i \leq n$}
\noLine
\UnaryInfC{$\rel, y_{i} \eq y_{j} \sar \la : {\geqslant} n \role . C, \cxti \ |\ 1 \leq i < j \leq n$}
\RightLabel{$\gtnr$}
\UnaryInfC{$\rel \sar \la : {\geqslant} n \role . C, \cxti$}
\DisplayProof
\end{center}
Assume, $\rel \sar \la : {\geqslant} n \role . C, \cxti$ is invalid relative to $\ont$. It follows that there exists an interpretation $\mathcal{I}$ and a label assignment $\lambda$ such that $\mathcal{I}, \lambda \amodels \rel$ and $\mathcal{I}, \lambda \not\emodels \la : {\geqslant} n \role . C, \cxti$. We note that if $n=0$, then $\lambda(x) \notin \map{({\geqslant} 0 \role . C)}$, which cannot be the case, and hence, we may suppose that $n > 0$. Then, we have that $\lambda(\la) \in \map{({\leqslant} (n{-}1) \role. C)}$. From the side condition on $\gtnr$, there are labels $y_1, \ldots, y_{n}$ such that $\eqclass{x}{\rel} \prpath{L} \eqclass{y_{i}}{\rel}$ with $L = \glang(\role)$, which, by \lem~\ref{lem:ppath-implies-R}, implies that $(\lambda(x),\lambda(y_{i})) \in \map{\role}$ for each $1 \leq i \leq n$. If $\lmap(y_{i}) \neq \lmap(y_{j})$ for each $1 \leq i < j \leq n$, then for some $1 \leq i \leq n$, we have that $\lambda(y_i) \notin \map{C}$; otherwise, we have $\lmap(y_{i}) = \lmap(y_{j})$ for some $1 \leq i < j \leq n$. The former cases show that at least one premise in the first set is invalid relative to $\ont$ and the latter case shows that at least one premise in the second set of premises is invalid relative to $\ont$.
\end{proof}

\begin{customlem}{\ref{prop:hp-admiss-wk}} The $\topr$ rule is provable in $\calc$, and the $\relab{x}{y}$, $\wkeq$, $\wkneq$, $\wk$, $\ctr$, and $\symneq$ rules are hp-admissible.
\end{customlem}

\begin{proof} Recall that $\top = A \dis \neg A$ for some fixed $A \in \cnames$. The proof below shows that $\topr$ provable in $\calc$.
\begin{center}
\AxiomC{}
\RightLabel{$\id$}
\UnaryInfC{$\rel \sar \cxti, x : A, x : \neg A$}
\RightLabel{$\disr$}
\UnaryInfC{$\rel \sar \cxti, x : A \dis \neg A$}
\DisplayProof
\end{center}

The hp-admissibility of $\relab{x}{y}$ is shown by induction on the height of the given proof. The base cases are trivial as any application of $\relab{x}{y}$ to an initial rule yields another instance of the initial rule. In the inductive step, with the exception of the $\allr$ and $\ltnr$ cases, all cases are resolved by invoking IH and then applying the rule. In the $\allr$ and $\ltnr$ cases, the label substituted into the sequent may be fresh in the $\allr$ or $\ltnr$ inference, requiring two applications of IH in order for the case to go through. To demonstrate this, we show how a problematic $\allr$ case is resolved and note that the $\ltnr$ case is similar. Suppose we have an instance of $\allr$ followed by $\relab{y}{z}$ as shown below left, where $y$ is fresh in $\allr$. The case is resolved as shown below right, where $\relab{v}{y}$ is applied in the first IH application with $v$ fresh due to the side condition of the rule, $\relab{y}{z}$ is then applied in the second IH application, and last, $\allr$ is applied.
\begin{center}
\begin{tabular}{c c}
\AxiomC{$\rel, \role(\la,\lb) \sar \lb : C, \lb : \gcilist{\ont}, \cxti$}
\RightLabel{$\allr$}
\UnaryInfC{$\rel \sar \la : \all C, \cxti$}
\RightLabel{$\relab{y}{z}$}
\UnaryInfC{$\rel(y/z) \sar \la : \all C, \cxti(y/z)$}
\DisplayProof

&

\AxiomC{$\rel, \role(\la,\lb) \sar \lb : C, \lb : \gcilist{\ont}, \cxti$}
\RightLabel{IH}
\UnaryInfC{$\rel, \role(\la,v) \sar v : C, v : \gcilist{\ont}, \cxti$}
\RightLabel{IH}
\UnaryInfC{$\rel(y/z), \role(\la,v) \sar v : C, v : \gcilist{\ont}, \cxti(y/z)$}
\RightLabel{$\allr$}
\UnaryInfC{$\rel(y/z) \sar \la : \all C, \cxti(y/z)$}
\DisplayProof
\end{tabular}
\end{center}

The hp-admissibility of $\wkeq$, $\wkneq$, and $\wk$ are shown by induction on the height of the given proof while making a case distinction on the last rule applied. The base cases are trivial as any application of either rule to $\id$ or $\ideq$ yields another instance of the rule, and with the exception of the $\gtnr$ case, every case of the inductive step may be resolved by applying IH followed by the corresponding rule. The $\gtnr$ case is trivial when showing the hp-admissibility of $\wkneq$ and $\wk$, however, an interesting case arises when showing the hp-admissibility of $\wkeq$. Suppose we have an application of $\gtnr$ as shown below, which weakens in an equality $y_{i} \eq y_{j}$ that is active in the $\gtnr$ application. Observe that the desired conclusion is obtained by taking the proof of the premise $\rel, y_{i} \eq y_{j} \sar \la : {\geqslant} n \role . C, \cxti$. All other cases where $\gtnr$ is followed by an application of $\wkeq$ are simple or resolved similarly.
\begin{center}
\AxiomC{$\rel \sar \lb_{i} : C, \la : {\geqslant} n \role . C, \cxti \  | \ 1 \leq i \leq n$}
\noLine
\UnaryInfC{$\rel, y_{i} \eq y_{j} \sar \la : {\geqslant} n \role . C, \cxti \ | \ 1 \leq i < j \leq n$}
\RightLabel{$\gtnr$}
\UnaryInfC{$\rel \sar \la : {\geqslant} n \role . C, \cxti$}
\RightLabel{$\wkeq$}
\UnaryInfC{$\rel, y_{i} \eq y_{j} \sar \la : {\geqslant} n \role . C, \cxti$}
\DisplayProof
\end{center}

The hp-admissibility of $\ctr$ is also shown by induction on the height of the given proof. We note that the hp-admissibility of $\ctr$ relies on the hp-admissibility of another contraction rule $\ctrr$, shown below left, which `fuses' two children nodes in a sequent. The hp-admissibility of $\ctr$ and $\ctrr$ is shown simultaneously by induction on the height of the given proof, though we focus on the $\ctr$ case as the $\ctrr$ case is similar. Returning back to the proof that $\ctr$ is hp-admissible, we note that the base cases are trivial since any application of $\ctr$ to $\id$ or $\ideq$ yields another instance of the rule. In the inductive step, if neither of the contraction formulae $x : C, x : C$ are principal in the conclusion of a rule application $(r)$, as shown in the example below right, then the case is resolved by applying IH (i.e. the $\ctr$ rule) and then the rule $(r)$. 
\begin{center}
\begin{tabular}{c c}
\AxiomC{$\rel, r(x,y), r(x,z) \sar \cxti$}
\RightLabel{$\ctrr$}
\UnaryInfC{$\rel(y/z), r(x,y) \sar \cxti(y/z)$}
\DisplayProof

&

\AxiomC{$\rel' \sar \cxti', x : C, x : C$}
\RightLabel{$(r)$}
\UnaryInfC{$\rel \sar \cxti, x : C, x : C$}
\RightLabel{$\ctr$}
\UnaryInfC{$\rel \sar \cxti, x : C$}
\DisplayProof
\end{tabular}
\end{center}
Therefore, let us suppose that one of the contraction formulae $x : C, x : C$ is principal. We consider the case where the last rule applied above $\ctr$ is the $\allr$ rule, as shown below left. To resolve the case, we apply the hp-invertibility of $\allr$, followed by the an application of $\ctrr$ which applies $(y/z)$, followed by a sufficient number of applications of IH for $\ctr$ to contract all of the displayed formulae in the consequent, and finally, an application of $\allr$, as shown below right.
\begin{center}
\begin{tabular}{c c}
\AxiomC{$\rel, r(x,y) \sar \cxti, y : C, y : \gcilist{\ont}, x : \all C$}
\RightLabel{$\allr$}
\UnaryInfC{$\rel \sar \cxti, x : \all C, x : \all C$}
\RightLabel{$\ctr$}
\UnaryInfC{$\rel \sar \cxti, x : \all C$}
\DisplayProof

&

\AxiomC{$\rel, r(x,y) \sar \cxti, y : C, y : \gcilist{\ont}, x : \all C$}
\RightLabel{Lem.~\ref{prop:hp-invert}}
\UnaryInfC{$\rel, r(x,y), r(x,z) \sar \cxti, y : C, y : \gcilist{\ont}, z : C, z : \gcilist{\ont}$}
\RightLabel{$\ctrr$}
\UnaryInfC{$\rel, r(x,y) \sar \cxti, y : C, y : \gcilist{\ont}, y : C, y : \gcilist{\ont}$}
\RightLabel{IH}
\UnaryInfC{$\rel, r(x,y) \sar \cxti, y : C, y : \gcilist{\ont}$}
\RightLabel{$\allr$}
\UnaryInfC{$\rel \sar \cxti, x : \all C$}
\DisplayProof
\end{tabular}
\end{center}
The remaining cases are solved in a similar fashion. Also, note that the proof of \lem~\ref{prop:hp-invert} does not rely on the hp-admissibility of $\ctr$ or $\ctrr$, and therefore, the above argument is not circular.

Last, we argue the hp-admissibility of $\symneq$ by induction on the height of the given proof. The only interesting case is the $\ideq$ case in the base case; the base case for $\id$ is trivial, and the inductive step follows in each case by applying IH followed by the rule. Suppose we have an instance of $\ideq$ as shown below left, where the side condition $\eqpath{x}{y}$ holds. Since $\eqpath{x}{y}$ holds \iffi $\eqpath{y}{x}$ by definition, we have that the application of $\ideq$ shown below right is a valid application of $\ideq$, thus showing $\symneq$ hp-admissible in this case.
\begin{center}
\begin{tabular}{c c}
\AxiomC{}
\RightLabel{$\ideq$}
\UnaryInfC{$\rel, x \noteq y, y \noteq x \sar \cxti$}
\RightLabel{$\symneq$}
\UnaryInfC{$\rel, y \noteq x \sar \cxti$}
\DisplayProof

&

\AxiomC{}
\RightLabel{$\ideq$}
\UnaryInfC{$\rel, y \noteq x \sar \cxti$}
\DisplayProof
\end{tabular}
\end{center}
The other case when $\symneq$ is applied to a non-principal inequality in $\ideq$ is trivial.
\end{proof}

\begin{customlem}{\ref{prop:hp-invert}}
All non-initial rules in $\calc$ are hp-invertible.
\end{customlem}

\begin{proof} The hp-invertibility of $\subeq$, $\existsr$, and $\gtnr$ follows from the fact that $\wk$ (and $\wkeq$ in the $\gtnr$ case) are hp-admissible (see Lemma~\ref{prop:hp-admiss-wk} above). The remaining cases are shown by induction on the height of the given proof. We only consider the $\ltnr$ case since all other cases are analogous.\\

\noindent
\textit{Base case}. Suppose we have instances of $\id$ and $\ideq$ as shown below.
\begin{center}
\begin{tabular}{c c}
\AxiomC{}
\RightLabel{$\idc$}
\UnaryInfC{$\rel \sar \la : A, \la : \neg A, \cxti, \lc : {\leqslant} n \role . C$}
\DisplayProof

\AxiomC{}
\RightLabel{$\ideq$}
\UnaryInfC{$\rel, x \noteq y \sar \cxti, \lc : {\leqslant} n \role . C$}
\DisplayProof
\end{tabular}
\end{center}
The instance of $\id$ shown below top and the instance of $\ideq$ shown below bottom resolve the base case. Note that we take $\rel' = \rel, \neqset{w_{0}, \ldots, w_{n}}, \role(\lc,w_{0}), \ldots, \role(\lc,w_{n})$.
\begin{center}
\AxiomC{}
\RightLabel{$\idc$}
\UnaryInfC{$\rel' \sar \la : A, \la : \neg A, \cxti, w_{0} : \negnnf{C}, w_{0} : \gcilist{\ont}, \ldots, w_{n} : \negnnf{C}, w_{n} : \gcilist{\ont}$}
\DisplayProof
\end{center}
\begin{center}
\AxiomC{}
\RightLabel{$\ideq$}
\UnaryInfC{$\rel', x \noteq y \sar \cxti, w_{0} : \negnnf{C}, w_{0} : \gcilist{\ont}, \ldots, w_{n} : \negnnf{C}, w_{n} : \gcilist{\ont}$}
\DisplayProof
\end{center}

\noindent
\textit{Inductive Step}. We only consider the $\subeq$ and $\ltnr$ cases as the remaining cases are similar.\\

\noindent
$\subeq$. Suppose we have an instance of $\subeq$ as shown below left. We can resolve the case as shown below right, where $\rel' = \rel, \neqset{w_{0}, \ldots, w_{n}}, \role(\lc,w_{0}), \ldots, \role(\lc,w_{n})$.
\begin{center}
\begin{tabular}{c c}
\AxiomC{$\rel \sar \la : L, \lb : L, \cxti, \lc : {\leqslant} n \role . C$}
\RightLabel{$\subeq$}
\UnaryInfC{$\rel \sar \la : L, \cxti, \lc : {\leqslant} n \role . C$}
\DisplayProof

&

\AxiomC{$\rel \sar \la : L, \lb : L, \cxti,  \lc : {\leqslant} n \role . C$}
\RightLabel{IH}
\UnaryInfC{$\rel' \sar \la : L, \lb : L, \cxti, w_{0} : \negnnf{C}, w_{0} : \gcilist{\ont}, \ldots, w_{n} : \negnnf{C}, w_{n} : \gcilist{\ont}$}
\RightLabel{$\subeq$}
\UnaryInfC{$\rel' \sar \la : L, \cxti, w_{0} : \negnnf{C}, w_{0} : \gcilist{\ont}, \ldots, w_{n} : \negnnf{C}, w_{n} : \gcilist{\ont}$}
\DisplayProof
\end{tabular}
\end{center}


\noindent
$\ltnr$. There are two cases to consider when the last rule applied is $\ltnr$. Either, the principal formula is the formula we want to invert, or it is not. In the first case, shown below, we simply take the proof of the premise, where $\rel' = \rel, \neqset{y_{0}, \ldots, y_{n}}, \role(\la,\lb_{0}), \ldots, \role(\la,\lb_{n})$.
\begin{center}
\AxiomC{$\rel' \sar  \lb_{0} : \negnnf{C}, \lb_{0} : \gcilist{\ont}, \ldots, \lb_{n} : \negnnf{C}, \lb_{n} : \gcilist{\ont}, \cxti$}
\RightLabel{$\ltnr$}
\UnaryInfC{$\rel \sar \la : {\leqslant} n \role . C, \cxti$}
\DisplayProof
\end{center}
In the second case, we simply apply IH and then the $\ltnr$ rule to resolve the case.
\end{proof}


\begin{customthm}{\ref{thm:completeness}}[Completeness]
If $\omodels \emptyset \sar \la : C$, then $\calc \Vdash \emptyset \sar \la : \gcilist{\ont}, \la : C$. 
\end{customthm}

\begin{proof} We describe a proof-search algorithm $\prove$ that takes a sequent $\sar \la : \gcilist, \la : C$ as input and attempts to construct a proof thereof. We assume that $\sar \la : \gcilist, \la : C$ is not provable in $\calc$ and show how to construct an interpretation $\inter = (\dom,\map{\cdot})$ and define a label assignment $\lmap$ such that $\inter, \lmap \not\omodels \emptyset \sar \la : \gcilist, \la : C$, meaning $\inter \vDash \ont$ and $\lmap(x) \not\in \map{C}$, that is, $\not\omodels \emptyset \sar \la : C$. Let us now begin our description of $\prove$.\\

\noindent
$\prove$. We take  $\sar \la : \gcilist{\ont}, \la : C$ as input and move to the step below.\\

\noindent
$\idc$ and $\ideq$. Let $\branch_{1}, \ldots, \branch_{n}$ be all branches of the pseudo-proof $\prf$ currently under construction. Let $\seq_{1}, \ldots, \seq_{n}$ be the top sequents of each branch, respectively. For each $1 \leq i \leq n$, if $\seq_{i}$ is an instance of $\id$ or $\ideq$, then apply $\idc$ and $\ideq$, respectively, bottom-up on $\branch_{i}$ and close that branch, i.e. halt $\prove$ on the branch $\branch_{i}$. If $\prove$ has halted on every branch, then return $\true$. Otherwise, if a sequent $\seq_{i}$ exists such that no rule from $\calc$ is bottom-up applicable to it, copy it above itself and continue to step $\subeq$ below.\\

\noindent
$\subeq$. Let $\branch_{1}, \ldots, \branch_{n}$ be all branches of the pseudo-proof $\prf$ currently under construction. Let $\seq_{1}, \ldots, \seq_{n}$ be the top sequents of each branch, respectively. We consider each branch and its top sequent in turn. Let us suppose we have already considered $\branch_{1}, \ldots, \branch_{i}$ so that $\branch_{i+1}$ is currently under consideration. Let the top sequent $\seq_{i+1}$ be of the form:
$$
\rel \sar \cxti, \la_{1} : L_{1}, \ldots, \la_{k} : L_{k}
$$
 with all labeled literals displayed. For each $1 \leq i \leq k$ and each $y \in \lab(\rel)$ such that $\eqpath{\la_{i}}{y}$, repeatedly apply the $\subeq$ rule bottom-up, extending $\branch_{i+1}$. After all branches have been processed in this way, we move onto the $\disr$ case below.\\

\noindent
$\disr$. Let $\branch_{1}, \ldots, \branch_{n}$ be all branches of the pseudo-proof $\prf$ currently under construction. Let $\seq_{1}, \ldots, \seq_{n}$ be the top sequents of each branch, respectively. We consider each branch and its top sequent in turn. Let us suppose we have already considered $\branch_{1}, \ldots, \branch_{i}$ so that $\branch_{i+1}$ is currently under consideration. Let the top sequent $\seq_{i+1}$ be of the form:
$$
\rel \sar \la_{1} : C_{1} \dis D_{1}, \ldots, \la_{k} : C_{k} \dis D_{k}, \cxti
$$
 with all disjunctive formulae displayed. We repeatedly apply the $\disr$ rule bottom-up, extending $\branch_{i+1}$ so that it now has a top sequent of the form:
 $$
\rel \sar \la_{1} : C_{1}, \la_{1} : D_{1}, \ldots, \la_{k} : C_{k}, \la_{k} : D_{k}, \cxti
$$
 After all branches have been processed in this way, we move onto the $\conr$ case below.\\

\noindent
$\conr$. Let $\branch_{1}, \ldots, \branch_{n}$ be all branches of the pseudo-proof $\prf$ currently under construction. Let $\seq_{1}, \ldots, \seq_{n}$ be the top sequents of each branch, respectively. We consider each branch and its top sequent in turn. Let us suppose we have already considered $\branch_{1}, \ldots, \branch_{i}$ so that $\branch_{i+1}$ is currently under consideration. Let the top sequent $\seq_{i+1}$ be of the form:
$$
\rel \sar \la_{1} : C_{1} \con D_{1}, \ldots, \la_{k} : C_{k} \con D_{k}, \cxti
$$
 with all conjunctive formulae displayed. We repeatedly apply the $\conr$ rule bottom-up, extending $\branch_{i+1}$ with $2^{k}$ new branches with each having a top sequent of the form:
 $$
\rel \sar \la_{1} : E_{1}, \ldots, \la_{k} : E_{k}, \cxti
$$
 where $E_{j} \in \{C_{j},D_{j}\}$ for $1 \leq j \leq k$. After all branches have been processed in this way, we move onto the $\existsr$ case below.\\

\noindent
$\existsr$. Let $\branch_{1}, \ldots, \branch_{n}$ be all branches of the pseudo-proof $\prf$ currently under construction. Let $\seq_{1}, \ldots, \seq_{n}$ be the top sequents of each branch, respectively. We consider each branch and its top sequent in turn. Let us suppose we have already considered $\branch_{1}, \ldots, \branch_{i}$ so that $\branch_{i+1}$ is currently under consideration. Let the top sequent $\seq_{i+1}$ be of the form:
$$
\rel \sar \la_{1} : \exists r_{1} . C_{1}, \ldots, \la_{k} : \exists r_{k} . C_{k}, \cxti
$$
 with all existential formulae displayed. For each $1 \leq j \leq k$ and every label $y \in \lab(\rel)$ such that $\eqclass{\la_{j}}{\rel} \prpath{L} \eqclass{y}{\rel}$ with $L = L_{\gram(\ont)}(r_{j})$, repeatedly apply the $\existsr$ rule bottom-up, extending $\branch_{i+1}$. After all branches have been processed in this way, we move onto the $\allr$ case below.\\
 
\noindent
$\allr$. Let $\branch_{1}, \ldots, \branch_{n}$ be all branches of the pseudo-proof $\prf$ currently under construction. Let $\seq_{1}, \ldots, \seq_{n}$ be the top sequents of each branch, respectively. We consider each branch and its top sequent in turn. Let us suppose we have already considered $\branch_{1}, \ldots, \branch_{i}$ so that $\branch_{i+1}$ is currently under consideration. Let the top sequent $\seq_{i+1}$ be of the form:
$$
\rel \sar \la_{1} : \forall r_{1} . C_{1}, \ldots, \la_{k} : \forall r_{k} . C_{k}, \cxti
$$
 with all universal formulae displayed. For each $1 \leq j \leq k$, repeatedly apply the $\allr$ rule bottom-up, extending $\branch_{i+1}$, and so the top sequent contains $r_{j}(x_{j},\lb)$ in the antecedent and $\lb : C_{j}, \lb : \gcilist{\ont}$ in the consequent, where $\lb$ is fresh for each $j$. After all branches have been processed in this way, we move onto the $\ltnr$ case below.\\

\noindent
$\ltnr$. Let $\branch_{1}, \ldots, \branch_{n}$ be all branches of the pseudo-proof $\prf$ currently under construction. Let $\seq_{1}, \ldots, \seq_{n}$ be the top sequents of each branch, respectively. We consider each branch and its top sequent in turn. Let us suppose we have already considered $\branch_{1}, \ldots, \branch_{i}$ so that $\branch_{i+1}$ is currently under consideration. Let the top sequent $\seq_{i+1}$ be of the form:
$$
\rel \sar \la_{1} : (\leqslant n_{1} \role_{1} . C_{1}), \ldots, \la_{k} : (\leqslant n_{k} \role_{k} . C_{k}), \cxti
$$
 with all qualified number restrictions of the form $\la_{j} : (\leqslant n_{j} \role_{j} . C_{j})$ displayed and where $1 \leq j \leq k$. We repeatedly apply the $\ltnr$ rule bottom-up, extending $\branch_{i+1}$ so that the top sequent contains $\neqset{y_{0}, \ldots, y_{n_{j}}}$ and $\role(\la_{j},y_{0}), \ldots, \role(\la_{j},y_{n_{j}})$ in the antecedent and $y_{0} : \negnnf{C}_{j}, \lb_{0} : \gcilist{\ont}, \ldots, y_{n_{j}} : \negnnf{C}_{j}, \lb_{n_{j}} : \gcilist{\ont}$ in the consequent with $y_{0}, \ldots, y_{n_{j}}$ fresh for each considered qualified number restriction of the above form. After all branches have been processed in this way, we move onto the $\gtnr$ case below.\\

\noindent
$\gtnr$. Let $\branch_{1}, \ldots, \branch_{n}$ be all branches of the pseudo-proof $\prf$ currently under construction. Let $\seq_{1}, \ldots, \seq_{n}$ be the top sequents of each branch, respectively. We consider each branch and its top sequent in turn. Let us suppose we have already considered $\branch_{1}, \ldots, \branch_{i}$ so that $\branch_{i+1}$ is currently under consideration. Let the top sequent $\seq_{i+1}$ be of the form:
$$
\rel \sar \la_{1} : (\geqslant n_{1} \role_{1} . C_{1}), \ldots, \la_{k} : (\geqslant n_{k} \role_{k} . C_{k}), \cxti
$$
 with all qualified number restrictions of the form $\la_{j} : (\geqslant n_{j} \role_{j} . C_{j})$ displayed and where $1 \leq j \leq k$. For each collection $\{y_{1}, \ldots, y_{n_{j}}\} \subseteq \lab(\seq_{i+1})$ of labels such that 
 $\eqclass{\la_{j}}{\rel} \prpath{L} \eqclass{y_{t}}{\rel}$ with $L = L_{\gram(\ont)}(r_{j})$ and $1 \leq t \leq n_{j}$, we apply the $\gtnr$ rule bottom-up, extending $\branch_{i+1}$. After all branches have been processed in this way, we cycle back to the $\idc$ and $\ideq$ case.\\
 
\noindent
This concludes the description of $\prove$.\\

 We know that $\prove$ cannot return $\true$ since then a proof of $\sar \la : \gcilist{\ont}, \la : C$ would exist, contrary to our assumption. Therefore, $\prove$ will not terminate, meaning that it constructs an infinite pseudo-proof $\prf$ in the form of an infinite tree. Since only finite branches occurs within this pseudo-proof, by K\"onig's lemma we know that an infinite branch of the following form exists in $\prf$:
$$
\branch = (\rel_{0} \sar \cxti_{0}), (\rel_{1} \sar \cxti_{1}), \ldots, (\rel_{n} \sar \cxti_{n}), \ldots
$$
 such that $\rel_{0} = \emptyset$ and $\cxti_{0} = \la : \gcilist{\ont}, \la : C$. Let us define $\rel^{*} = \bigcup_{i \in \mathbb{N}} \rel_{i}$ and $\cxti^{*} = \bigcup_{i \in \mathbb{N}} \cxti_{i}$. We use $\branch$ to construct an interpretation $\inter = (\dom,\map{\cdot})$ and define a label assigment $\lmap$ such that $\inter \vDash \ont$, $\inter, \lmap \amodels \rel^{*}$, but $\inter, \lmap \not\emodels \cxti^{*}$, meaning $\not\omodels \emptyset \sar \la : C$. We define $\dom = \{[y] \ | \ y \in \lab(\branch) \}$, where we use $[y] = [y]_{\rel^{*}}$ for simplicity. In other words, $\dom$ contains all equivalence classes modulo the $=^{*}_{\rel^{*}}$ relation on labels occurring in the sequents of the branch $\branch$. We define $\lmap$ and $\map{\cdot}$ accordingly:
\begin{itemize}

\item $\map{\lb} = [\lb]$ for $\lb \in \lab(\branch)$;

\item $[\lb] \in \map{A}$ \iffi $\lb : \neg A \in \cxti^{*}$;

\item $([\lb],[\lc]) \in \map{\role}$ \iffi either (1) there exist $w \in [\lb]$ and $u \in [\lc]$ such that $\role(w,u) \in \rel^{*}$, or (2) $\role_{1} \circ \cdots \circ \role_{n}\imp \role  \in \ont$ and $([\la]_{i-1}, [\la]_{i}) \in \map{\role}_{i}$ for $1 \leq i \leq n$ with $\lb \in [\la_{0}]$ and $\lc \in [\la_{n}]$. 

\end{itemize}

Based on the definition above and the definition of an $\roles$-system $\gram(\ont)$, we note that if $([y],[z]) \in \map{\role}$, then there exist roles $r_{1}, \ldots, r_{n}$ and $[z_{1}], \ldots, [z_{n-1}]$ such that
$$
\prgr{\rel^{*}} \vDash [y] \prpath{r_{1}} [z_{1}] \prpath{r_{2}} \cdots [z_{n-1}] \prpath{r_{n}} [z]
$$
and $r_{1} \cdots r_{n} \in L_{\gram(\ont)}(r)$. It is straightforward to show that $\inter$ satisfies all RIAs $\role_{1} \circ \cdots \circ \role_{n} \imp \role$ in $\ont$. We now argue that $\inter, \lmap \amodels \rel^{*}$. If $r(y,z) \in \rel^{*}$, then $([\lb],[\lc]) \in \map{\role}$ holds by the definition above, showing that $(\lmap(\lb),\lmap(\lc)) \in \map{\role}$. If $y \eq z \in \rel^{*}$, then by definition $[y] = [z]$, showing $\lmap(y) = \lmap(z)$. Also, observe that if $y \noteq z \in \rel^{*}$, then it cannot be the case that $[y] = [z]$ since then the $\ideq$ rule would be applied in $\branch$, implying that $\branch$ is finite, which is a contradiction as $\branch$ is infinite; hence, $\lmap(y) \neq \lmap(z)$. 
 We will now show that $\inter, \lmap \not\emodels \cxti^{*}$, and afterward, we will argue that $\inter$ satisfies all GCIs in $\ont$, thus establishing that $\inter \vDash \ont$. We argue by induction on the complexity of $D$ that if $y : D \in \cxti^{*}$, then $\lmap(y) \not\in \map{D}$.

\begin{description}

\item[$y : A \in \cxti^{*}.$] If $y : A \in \cxti^{*}$, then for every $z \in [y]$, $z : \neg A \not\in \cxti^{*}$ since otherwise $\subeq$ would have been applied, eventually followed by $\id$ and $\branch$ would be finite, contrary to our assumption. Therefore, $y : \neg A \not\in \cxti^{*}$, implying that $\lmap(y) \notin \map{A}$ by the definition of $\inter$.\\

\item[$y : \neg A \in \cxti^{*}.$] Then, $\lmap(y) \in \map{A}$, by the definition of $\inter$.\\

\item[$y : E \dis F \in \cxti^{*}.$] Then, eventually the $\disr$ rule will be applied in $\prove$, meaning $y : E, y : F \in \cxti^{*}$, showing that $\lmap(y) \in \map{E}$ and $\lmap(y) \in \map{F}$ by IH. Therefore, $\lmap(y) \in \map{(E \dis F)}$.\\

\item[$y : E \con F \in \cxti^{*}.$] If $y : E \con F \in \cxti$, then eventually the $\conr$ rule will be applied in $\prove$, meaning either $y : E \in \cxti$ or $y : D \in \cxti$. Hence, either $\lmap(y) \in \map{E}$ or $\lmap(y) \in \map{F}$ by IH, implying $\lmap(y) \in \map{(E \con F)}$.\\

\item[$\lb : \some E \in \cxti^{*}.$] Suppose $(\lmap(y),\lmap(z)) \in \map{r}$, i.e. $([y],[z]) \in \map{r}$. It follows that $[\lb] \prpath{L} [z]$ with $L = L_{\gram(\ont)}(r)$, meaning at some step in $\branch$, we have that $[\lb]_{\rel} \prpath{L} [z]_{\rel}$ with $L = L_{\gram(\ont)}(r)$. Hence, the $\existsr$ rule will be bottom-up applied, ensuring that $\lc : E \in \cxti^{*}$. By IH, we have that $\lmap(z) \notin \map{E}$, so since $z$ was assumed arbitrary, we have that $\lmap(y) \notin \map{(\some E)}$.\\

\item[$\lb : \all E \in \cxti^{*}.$] If $\lb : \all E \in \cxti^{*}$, then eventually the $\allr$ rule will be bottom-up applied in $\prove$, ensuring that for some label $z$, $r(y,z) \in \rel^{*}$ and $\lc : E \in \cxti^{*}$. By the definition of $\map{r}$, $([y],[z]) \in \map{r}$, and by IH $\lmap(z) \notin \map{E}$, thus $\lmap(y) \notin \map{\all E}$.\\

\item[$\lb : {\leqslant} n r . E \in \cxti^{*}.$] If $\lb: {\leqslant} n r . E \in \cxti$, then eventually the $\ltnr$ rule will be applied bottom-up in $\prove$. It follows that $\neqset{z_{0}, \ldots, z_{n}}, r(y,\lc_{0}), \ldots, r(y,\lc_{n}) \in \rel^{*}$ and $\lb_{0} : \negnnf{E}, \ldots, \lb_{n} : \negnnf{E} \in \cxti^{*}$. Therefore, there exist at least $n+1$ distinct elements in $\dom$, namely $[y_{0}], \ldots, [y_{n}]$, such that for $0 \leq i \leq n$, $([y],[\lc_{i}]) \in \map{r}$ and where $\lmap(z_{i}) \notin \map{E_{i}}$ by IH. Hence, $\lmap(y) \notin \map{({\leqslant} n r . E)}$.\\

\item[$y : {\geqslant} n r . E \in \cxti^{*}.$] Assume that distinct $[\lc_{1}], \ldots, [\lc_{n}] \in \dom$ exist such that for $1 \leq i \leq n$, $([\lb],[\lc_{i}]) \in \map{\role}$. It follows that for $1 \leq i \leq n$, $[y] \prpath{L} [z_{i}]$ with $L = \glang(\role)$ and $z_{1}, \ldots, z_{n}$ pairwise, distinct labels. Hence, at some step in $\branch$, we have that $\eqclass{y}{\rel} \prpath{L} \eqclass{z_{i}}{\rel}$ with $L = \glang(\role)$, and so, the $\gtnr$ rule will be applied bottom-up, ensuring that $z_{i} : E \in \cxti^{*}$ for some $1 \leq i \leq n$. By IH, $\lmap(z_{i}) \notin \map{E}$. Since the elements $[\lc_{1}], \ldots, [\lc_{n}] \in \dom$ were assumed to be arbitrary and distinct, this shows that for any $n$ elements that $\lmap(\lb)$ relates to via $\map{\role}$, at least one must not be an element of $\map{E}$, i.e. $\lmap(\lb) \notin \map{({\geqslant} n r . E)}$.

\end{description}
Let us now argue that $\inter \models \ont$ by arguing that all GCIs in $\ont$ are satisfied on $\inter$ (note that all RIAs are satisfied on $\inter$ as stated above). Observe that our input is of the form $\sar \la : \gcilist{\ont}, \la : C$ and every time a fresh label is added by the $\allr$ or $\ltnr$ step of $\prove$, the concepts in $\gcilist{\ont}$ are introduced at that label, meaning such concepts will occur at every label in $\lab(\rel^{*},\cxti^{*})$. Therefore, by the argument above, we know that for every GCI $\top \imp E \in \ont$ and $\lmap(y) = [y] \in \dom$, $\lmap(y) \notin \map{\negnnf{E}}$, showing that $\lmap(y) \in \map{E}$, and hence, every GCI is satisfied on $\inter$. As $\la : C \in \cxti^{*}$, all of the above implies that $\omodels \emptyset \sar \la : C$, completing the proof.
\end{proof}

%% file: app-2.tex
\begin{customlem}{\ref{lem:main-interpolation-lemma-1}}
Let $\ont = \ont_{1} \cup \ont_{2}$ be an ontology 
and suppose that $\rel, \Phi, \Psi \sar \gcilist{\ont_{a}}, \cxti, \cxtii, \gcilist{\ont_{b}}$ has a proof $\prf$ in $\calc$ with $\Phi \cap \Psi = \emptyset$. Then, $\prf$ can be transformed into a proof in $\icalc$ of $\rel; \Phi \splt \Psi \sar \gcilist{\ont_{a}}, \cxti \splt \cxtii, \gcilist{\ont_{b}} \sep \itp$ such that:
\begin{itemize}

\item[(1)] If $x \eq y$ occurs in $\itp$, then $x \noteq y \in \Phi$ ;

\item[(2)] If $x \noteq y$ occurs in $\itp$, then $x \noteq y \in \Psi$;

\item[(3)] $\lab(\itp) \subseteq \lab(\rel, \Phi, \Psi, \gcilist{\ont_{a}}, \cxti, \cxtii, \gcilist{\ont_{b}})$; 

\item[(4)] $\cpt{\itp} \subseteq \cpt{\ont_{a}, \cxti} \cap \cpt{\cxtii, \ont_{b}}$.
 
\end{itemize}
\end{customlem}

\begin{proof} By induction on the height of the proof. The base cases are trivial, so we focus on the inductive step. We argue one of the $\ltnr$ cases as the remaining cases are simple or similar. Suppose we have a proof in $\calc$ ending with an application of $\ltnr$, as shown below, where $\rel' = \rel, \role(\la,\lb_{0}), \ldots, \role(\la,\lb_{n})$ and let $c \in \{a,b\}$.
\begin{center}
\AxiomC{$\rel', \Phi,\neqset{y_{0}, \ldots, y_{n}}, \Psi \sar \gcilist{\ont_{a}}, \cxti, \lb_{0} : \negnnf{C}, \ldots, \lb_{n} : \negnnf{C}, \cxtii, \gcilist{\ont_{b}}$}
\RightLabel{$\ltnr$}
\UnaryInfC{$\rel, \Phi, \Psi \sar \gcilist{\ont_{a}}', \cxti, \la : {\leqslant} n \role . C, \cxtii, \gcilist{\ont_{b}}'$}
\DisplayProof
\end{center}
By IH, the premise of the inference shown below has a proof in $\icalc$, where $\gcilist{\ont_{c}} = \gcilist{\ont_{c}}', \lb_{0} : \gcilist{\ont_{c}}, \ldots, \lb_{n} : \gcilist{\ont_{c}}$ for $c \in \{a,b\}$. Since properties (1) and (2) hold for the premise below, and $\Phi \cap (\neqset{y_{0}, \ldots, y_{n}}, \Psi) = \emptyset$ by assumption, we know $y_{0}, \ldots, y_{n} \not\in \lab(\Phi, \Psi)$ since otherwise the side condition of the $\ltnr$ application above would be violated. Hence, $\itp$ will be in the form dictated by \dfn~\ref{def:special-interpolants}, meaning we obtain a proof of the desired conclusion in $\icalc$ by a single application of $\ltnr$, as shown below.
\begin{center}
\AxiomC{$\rel'; \Phi \splt \neqset{y_{0}, \ldots, y_{n}}, \Psi \sar \gcilist{\ont_{a}}, \cxti \splt \lb_{0} : \negnnf{C}, \ldots, \lb_{n} : \negnnf{C}, \cxtii, \gcilist{\ont_{b}} \sep \itp$}
\RightLabel{$\ltnr$}
\UnaryInfC{$\rel; \Phi \splt \Psi \sar \gcilist{\ont_{a}}', \cxti \splt \la : {\leqslant} n \role . C, \cxtii, \gcilist{\ont_{b}}' \sep \leqnr. \itp$}
\DisplayProof
\end{center}
$\lab(\itp) \subseteq \lab(\rel, \Phi, \Psi, \gcilist{\ont_{a}}', \cxti, \cxtii, \gcilist{\ont_{b}}') {\cup} \{y_{0}, \ldots, y_{n}\}$ and $\cpt{\itp} {\subseteq} \cpt{\ont_{a}, \cxti} {\cap} \cpt{y_{0} {:} \negnnf{C}_{0}, \ldots, y_{n} {:} \negnnf{C}_{n}, \cxtii, \ont_{b}}$ follow from IH with the latter also following from the fact that $\cpt{\gcilist{\ont_{c}}} \subseteq \cpt{\ont_{c}}$. (1) and (2) easily hold, and as shown below, property (3) follows from the former fact:
\begin{align*}
\lab(\leqnr . \itp) & =  \lab(\itp) \setminus \{y_{0}, \ldots, y_{n}\}\\
& \subseteq  \lab(\rel, \Phi, \Psi, \gcilist{\ont_{a}}', \cxti, \cxtii, \gcilist{\ont_{b}}')\\
& \subseteq \lab(\rel, \Phi, \Psi, \gcilist{\ont_{a}}', \cxti, \la : {\leqslant} {n} \role . C, \cxtii, \gcilist{\ont_{b}}')
\end{align*}
while property (4) follows from the latter fact:
\begin{align*}
& \cpt{\leqnr . \itp} =  \ \cpt{\itp} \\
& \subseteq \cpt{\ont_{a}, \cxti} \cap \cpt{y_{0} : \negnnf{C}_{0}, \ldots, y_{n} : \negnnf{C}_{n}, \cxtii, \ont_{b}}\\
& = \cpt{\ont_{a}, \cxti} \cap \cpt{\la : {\leqslant} n \role . C, \cxtii, \ont_{b}}\qedhere
\end{align*}
\end{proof}

\begin{customlem}{\ref{lem:orth-lemma}}
If $(\cxtii \sar \cxtiii) \in \orth{\orth{\itp}}$, then there exists a $(\rel \sar \cxti) \in \itp$ such that $\rel \subseteq \cxtii$ and $\cxti \subseteq \cxtiii$. 
\end{customlem}

\begin{proof} Let $(\cxtii \sar \cxtiii) \in \orth{\orth{\itp}}$ and suppose for a contradiction that no $(\rel \sar \cxti) \in \itp$ exists such that $\rel \subseteq \cxtii$ and $\cxti \subseteq \cxtiii$. Then, for each $(\rel \sar \cxti) \in \itp$, either $\rel \not\subseteq \cxtii$ or $\cxti \not\subseteq \cxtiii$. Let $\itp = \{(\rel_{1} \sar \cxti_{1}), \ldots, (\rel_{n} \sar \cxti_{n})\}$ and define $\rel' \sar \cxti'$ such that for each $1 \leq i \leq n$, one and only one of the following holds: (1) $x \eq y \in \rel'$ and $x \noteq y \in \rel_{i} \setminus \cxtii$, (2) $x \noteq y \in \rel'$ and $x \eq y \in \rel_{i} \setminus \cxtii$, or (3) $x : \negnnf{C} \in \cxti'$ and $x : C \in \cxti_{i} \setminus \cxtiii$. Observe that $\cxtii \cap \rel' =\emptyset$ and $\cxtiii \cap \cxti' =\emptyset$ by construction. However, since $\rel' \sar \cxti' \in \orth{\itp}$ by \dfn~\ref{def:sequent-interpolant}, we have that either $\cxtii \cap \rel' \neq \emptyset$ or $\cxtiii \cap \cxti' \neq \emptyset$, which gives a contradiction and proves the lemma.
\end{proof}

\begin{customlem}{\ref{lem:interpolation-lemma}}
If $\icalc \Vdash \rel; \Phi \splt \Psi \sar \cxti \splt \cxtii \sep \itp$, then 
\begin{enumerate}

\item For each $(\rel' \sar \cxtiii_{i}) \in \itp$, $\calc \Vdash \rel, \rel', \Phi \sar \cxti, \cxtiii_{i}$; 

\item For each $(\rel' \sar \cxtiv_{i}) \in \orth{\itp}$, $\calc \Vdash \rel, \rel', \Psi \sar \cxtiv_{i}, \cxtii$. 

\end{enumerate}
\end{customlem}

\begin{proof} We prove both claims simultaneously by induction on the height of the proof and make a case distinction on the last rule applied.\\ 

\noindent
\emph{Base case.} Suppose we have a proof of height $1$, i.e. an instance of an initial rule. Let us first consider the $\idcIii$ case and then we will consider the $\ideqI$ case, noting that the $\idcIi$ case is similar.
\begin{center}
\AxiomC{}
\RightLabel{$\idcIii$}
\UnaryInfC{$\rel; \Phi \splt \Psi \sar \cxti \splt \la : A, \la : \neg A, \cxtii \sep \{(\ \sar \la : \top)\}$}
\DisplayProof
\end{center}
Claim 1 is resolved as shown below left, and relies on \lem~\ref{prop:hp-admiss-wk}, whereas claim 2 is resolved by applying the $\idc$ rule. Note that we use the interpolant $\orth{\itp} = \{(\ \vdash \la : \bot)\}$ in claim 2 (see \dfn~\ref{def:sequent-interpolant} above) and $\cxtii' = \cxtii, \la : A, \la : \neg A$.
\begin{center}
\begin{tabular}{c c}
\AxiomC{}
\RightLabel{$\topr$}
\UnaryInfC{$\rel, \Phi \sar \cxti, \la : \top$}
\DisplayProof

&

\AxiomC{}
\RightLabel{$\idc$}
\UnaryInfC{$\rel, \Psi \sar \cxtii', \la : \bot$}
\DisplayProof
\end{tabular}
\end{center}
\noindent
For the $\ideqI$ case, suppose we have a proof consisting of a single application of the $\ideqI$ rule, as shown below.
\begin{center}
\AxiomC{}
\RightLabel{$\ideqI$}
\UnaryInfC{$\rel; \Phi \splt \Psi, x \noteq y  \sar \cxti \splt \cxtii \sep \{ (x \noteq y \sar \ ) \}$}
\DisplayProof
\end{center}
By the side condition on the rule, we know that $\eqpath{x}{y}$ holds for $\ideqI$ above. Therefore, the application of $\ideq$ shown below left is warranted as is the application of $\ideq$ shown below right, thus resolving the case.
\begin{center}
\begin{tabular}{c c}
\AxiomC{}
\RightLabel{$\ideqI$}
\UnaryInfC{$\rel, \Phi, x \noteq y  \sar \cxti$}
\DisplayProof

&

\AxiomC{}
\RightLabel{$\ideqI$}
\UnaryInfC{$\rel, \Psi, x \noteq y, x \eq y  \sar \cxtii$}
\DisplayProof
\end{tabular}
\end{center}

\noindent
\emph{Inductive step.} For the inductive step, we consider the $\orthru$, $\conrI$, $\allrI$, and $\ltnrI$ cases as the remaining cases are similar.\\

\noindent
$\orthru$. Let our proof in $\icalc$ end with an $\orthru$ application:
\begin{center}
\AxiomC{$\rel; \Phi \splt \Psi \sar \cxti \splt \cxtii \sep \itp$}
\RightLabel{$\orthru$}
\UnaryInfC{$\rel; \Psi \spltrev \Phi \sar \cxtii \spltrev \cxti \sep \orth{\itp}$}
\DisplayProof
\end{center}
Observe that if $(\rel' \sar \cxtiii_{i}) \in \orth{\itp}$, then $\rel, \Psi, \rel' \sar \cxtii, \cxtiii_{i}$ is provable in $\calc$ by IH, which demonstrates claim (1). To prove claim (2), let $(\rel' \sar \cxtiii_{i}) \in \orth{\orth{\itp}}$. Then, by Lemma~\ref{lem:orth-lemma}, there exists a $(\cxtii' \sar \cxtiii_{i}') \in \itp$ such that $\cxtii' \subseteq \rel'$ and $\cxtiii_{i}' \subseteq \cxtiii_{i}$. By IH, $\calc \Vdash \rel, \Phi, \cxtii' \sar \cxtiii_{i}', \cxti$, implying $\calc \Vdash \rel, \Phi, \rel' \sar \cxtiii_{i}, \cxti$ by the hp-admissibility of $\wkeq$, $\wkneq$, and $\wk$ (Lemma~\ref{prop:hp-admiss-wk}).\\

\noindent
$\conrI$. Suppose we have a proof in $\icalc$ ending with an application of $\conrI$, as shown below.
\begin{center}
\AxiomC{$\rel; \Phi \splt \Psi \sar \cxti \splt \la : C, \cxtii \sep \itp_{1}$}
\AxiomC{$\rel; \Phi \splt \Psi \sar \cxti \splt \la : D, \cxtii \sep \itp_{2}$}
\RightLabel{$\conrI$}
\BinaryInfC{$\rel; \Phi \splt \Psi \sar \cxti \splt \la : C \dis D, \cxtii \sep \itp_{1} \cup \itp_{2}$}
\DisplayProof
\end{center}
We prove claim (1) first. Let $(\rel' \sar \cxtiii) \in \itp_{1} \cup \itp_{2}$. Regardless of if $(\rel' \sar \cxtiii) \in \itp_{1}$ or $(\rel' \sar \cxtiii) \in \itp_{2}$, we have a proof of $\rel, \Phi, \rel' \sar \cxti, \cxtiii$ by IH, which proves the claim. To prove claim (2), suppose that $(\rel_{1}, \rel_{2} \sar \cxtiii_{1}, \cxtiii_{2}) \in \orth{\itp_{1} \cup \itp_{2}}$ such that $(\rel_{1} \sar \cxtiii_{1}) \in \orth{\itp}_{1}$ and $(\rel_{2} \sar  \cxtiii_{2}) \in \orth{\itp}_{2}$. By IH, the top two sequents below are provable in $\calc$. We then apply the hp-admissibility of $\wkeq$, $\wkneq$, and $\wk$ (\lem~\ref{prop:hp-admiss-wk}) a sufficient number of times (indicated by the asterisk) to ensure the contexts match. Last, we apply the $\conr$ to obtain the desired conclusion.
\begin{center}
\AxiomC{$\rel, \rel_{1}, \Psi \sar \la : C, \cxtii, \cxtiii_{1}$}
\RightLabel{$\wkeq^{*}, \wkneq^{*}, \wk^{*}$}
\UnaryInfC{$\rel, \rel_{1}, \rel_{2}, \Psi \sar \la : C, \cxtii, \cxtiii_{1}, \cxtiii_{2}$}

\AxiomC{$\rel, \rel_{2}, \Psi \sar \la : D, \cxtii, \cxtiii_{2}$}
\RightLabel{$\wkeq^{*}, \wkneq^{*}, \wk^{*}$}
\UnaryInfC{$\rel, \rel_{1}, \rel_{2}, \Psi \sar \la : D, \cxtii, \cxtiii_{1}, \cxtiii_{2}$}

\RightLabel{$\conr$}
\BinaryInfC{$\rel, \rel_{1}, \rel_{2} \Psi \sar \la : C \dis D, \cxtii, \cxtiii_{1}, \cxtiii_{2}$}
\DisplayProof
\end{center}

\noindent
$\allrI$. Suppose our proof ends with an $\allrI$ application:
\begin{center}
\AxiomC{$\rel, \role(\la,\lb); \Phi \splt \Psi  \sar \cxti \splt \lb : C, \lb : \gcilist{\ont}, \cxtii \sep \itp$}
\RightLabel{$\allrI$}
\UnaryInfC{$\rel; \Phi \splt \Psi \sar \cxti \splt \la : \all C, \cxtii \sep \all \itp$}
\DisplayProof
\end{center}
 We first argue claim (1). Let $(\rel' \sar \cxtiii, y : C_{1}, \ldots, y : C_{n}) \in \itp$ such that all concepts labeled with $y$ are displayed and $y \not\in \lab(\rel')$ by assumption, and let $\clist = C_{1}, \ldots, C_{n}$. By IH, we obtain a proof in $\calc$ ending as shown below: 
\begin{center}
\AxiomC{$\rel, \rel', r(x,y) \sar \cxti, \cxtiii, y : C_{1}, \ldots, y : C_{n}$}
\RightLabel{$\disr \times (n{-}1)$}
\UnaryInfC{$\rel, \rel', \role(\la,\lb) \sar \cxti, \cxtiii, y : \bigsqcup \clist$}
\RightLabel{$\allr$}
\UnaryInfC{$\rel, \rel' \sar \cxti, \cxtiii, \la : \all \bigsqcup \clist$}
\DisplayProof
\end{center}
We now argue claim (2). Observe that for any $(\rel' \sar \cxtiii) \in \orth{\itp}$, the multiset $\cxtiii$ contains zero or more labeled concepts of the form $x : \some (\negnnf{C}_{1} \con \cdots \con \negnnf{C}_{n})$. We suppose $\cxtiii$ contains one such formula as the other cases are analogous, and let $\cxtiii = \cxtiii', x : \some (\negnnf{C}_{1} \con \cdots \con \negnnf{C}_{n})$. The case is resolved as shown below, where $\conr^{*}$ denotes $n{-}1$ applications of $\conr$ between the $n$ premises obtained from IH and $\cxtii' = \lb : C, \lb : \gcilist{\ont}, \cxtii$.
\begin{center}
\AxiomC{$\rel, \rel', \role(\la,\lb) \sar \cxtii', \cxtiii', y : \negnnf{C}_{i} \ | \ 1 \leq i \leq n$}
\RightLabel{$\conr^{*}$}
\UnaryInfC{$\rel, \rel', \role(\la,\lb) \sar \cxtii', \cxtiii', y : \bigsqcap \negnnf{\clist}$}
\RightLabel{$\wk$}
\UnaryInfC{$\rel, \rel', \role(\la,\lb) \sar  \la : \some \bigsqcap \negnnf{\clist}, \cxtii', \cxtiii', y : \bigsqcap \negnnf{\clist}$}
\RightLabel{$\existsr$}
\UnaryInfC{$\rel, \rel', \role(\la,\lb) \sar  \la : \some \bigsqcap \negnnf{\clist}, \cxtii', \cxtiii'$}
\RightLabel{$\allr$}
\UnaryInfC{$\rel, \rel' \sar  \la : \some \bigsqcap \negnnf{\clist}, \la : \all C, \cxtii, \cxtiii'$}
\DisplayProof
\end{center}

\noindent
$\ltnrI$. Let us suppose that we have a proof in $\icalc$ ending with an application of $\ltnrI$ as shown below such that $\rel' = \rel,  \role(\la,\lb_{0}), \ldots, \role(\la,\lb_{n})$ and where $\cxtii' = \lb_{0} : \negnnf{C}, \lb_{0} : \gcilist{\ont}, \ldots, \lb_{n} : \negnnf{C}, \lb_{n} : \gcilist{\ont}, \cxtii$.
\begin{center}
\AxiomC{$\rel'; \Phi \splt \Psi, \neqset{y_{0}, \ldots, y_{n}} \sar \cxti \splt \cxtii' \sep \itp$}
\RightLabel{$\ltnrI$}
\UnaryInfC{$\rel; \Phi \splt \Psi \sar \cxti \splt \la : (\leqslant n \role . C), \cxtii \sep \leqnr. \itp$}
\DisplayProof
\end{center}
We argue claim (1) first and suppose we have the following: $(\rel_{1}, \rel_{2} \sar \cxtiii, y_{0} : \clist_{0}, \ldots, y_{n} : \clist_{n}) \in \itp$ such that $\rel_{2} \subseteq \neqset{y_{0}, \ldots, y_{n}}$, $\lab(\rel_{1}) \cap \{y_{0}, \ldots, y_{n}\} = \emptyset$, and we let $\clist = \clist_{0}, \ldots, \clist_{n}$. By IH, the top sequent shown below has a proof in $\calc$. We invoke \lem~\ref{prop:hp-admiss-wk} and apply the hp-admissible $\wk$ and $\wkneq$ rules  a sufficient number of times to derive the second sequent from the first, and then apply the $\disr$ rule a sufficient number of times to obtain the third sequent from the second, using an asterisk $*$ to indicate these sequential rule applications.
\begin{center}
\AxiomC{$\rel', \Phi, \rel_{1}, \rel_{2} \sar \cxti, \cxtiii, y_{0} : \clist_{0}, \ldots, y_{n} : \clist_{n}$}
\RightLabel{$\wk^{*}, \wkneq^{*}$}
\UnaryInfC{$\rel', \Phi, \rel_{1}, \neqset{y_{0}, \ldots, y_{n}} \sar \cxti, \cxtiii, y_{0} : \clist, \ldots, y_{n} : \clist$}
\RightLabel{$\disr^{*}$}
\UnaryInfC{$\rel', \Phi, \rel_{1}, \neqset{y_{0}, \ldots, y_{n}} \sar \cxti, \cxtiii, y_{0} : \bigsqcup \clist, \ldots, y_{n} : \bigsqcup \clist$}
\RightLabel{$\ltnr$}
\UnaryInfC{$\rel, \Phi, \rel_{1} \sar \cxti, \cxtiii, x : \leqnr . \negnnf{\bigsqcup \clist}$}
\DisplayProof
\end{center}
Let us now argue claim (2). Observe if $(\rel'' \vdash \cxtiii) \in \orth{\leqnr. \itp}$, then $\cxtiii$ contains zero or more labeled concepts of the form $x : \geqnir . \bigsqcup \clist$. We suppose $\cxtiii$ contains one such formula and remark that the remaining cases are similar. Therefore, $\cxtiii = \cxtiii', x : \geqnir . \bigsqcup \clist$. We let $\clist_{i} = C_{i,1}, \ldots, C_{i,k_{i}}$ with $0 \leq i \leq n$ and $\clist = \clist_{0}, \ldots, \clist_{n}$. By IH, for each $0 \leq i \leq n$, each top sequent shown below is provable in $\calc$. We apply the $\conr$ rule $k_{i}{-}1$ times, indicated by $\conr^{*}$,  to derive the second sequent from the first. 
\begin{center}
\AxiomC{$\rel', \Psi, \neqset{y_{0}, \ldots, y_{n}}, \rel'' \sar \cxtii', \cxtiii', y_{i} : \negnnf{C}_{i,j} \ | \ 1 \leq j \leq k_{i}$}
\RightLabel{$\conr^{*}$}
\UnaryInfC{$\rel', \Psi, \neqset{y_{0}, \ldots, y_{n}}, \rel'' \sar \cxtii', \cxtiii', y_{i} : \bigsqcap \negnnf{\clist_{i}}$}
\DisplayProof
\end{center}
We denote the concluding sequent in the proof above as $\seq_{i}'$. Let us pick an $1 \leq i \leq n$ and let $1 \leq \ell \neq i \leq n$. We then take $\seq_{\ell}'$ and apply the hp-admissible rules $\relab{z}{y_{\ell}}$, $\relab{y_{\ell}}{y_{i}}$, and $\relab{y_{i}}{z}$ with $z$ fresh to obtain the sequent $\seq_{\ell}''$, shown below top, where the labels $y_{i}$ and $y_{\ell}$ have been `swapped'. Observe that some inequalities of the form $y_{t} \noteq y_{s}$ in $\neqset{y_{0}, \ldots, y_{n}}$ may have `flipped' to $y_{s} \noteq y_{t}$, yielding the set $\neqsetii{y_{0}, \ldots, y_{n}}$. By applying $\wkneq$ a sufficient number of times, one can weaken in all inequalities from $\neqset{y_{0}, \ldots, y_{n}}$ not occurring in $\neqsetii{y_{0}, \ldots, y_{n}}$, and then apply $\symneq$ a sufficient number of times so that only $\neqset{y_{0}, \ldots, y_{n}}$ occurs in the antecedent. 
\begin{center}
\AxiomC{$\rel', \Psi, \neqsetii{y_{0}, \ldots, y_{n}}, \rel'' \sar \cxtii', \cxtiii', y_{i} : \bigsqcap \negnnf{\clist_{\ell}}$}
\RightLabel{$\wkneq^{*}$}
\UnaryInfC{$\rel', \Psi, \neqsetii{y_{0}, \ldots, y_{n}}, \neqset{y_{0}, \ldots, y_{n}}, \rel'' \sar \cxtii', \cxtiii', y_{i} : \bigsqcap \negnnf{\clist_{\ell}}$}
\RightLabel{$\symneq^{*}$}
\UnaryInfC{$\rel', \Psi, \neqset{y_{0}, \ldots, y_{n}}, \rel'' \sar \cxtii', \cxtiii', y_{i} : \bigsqcap \negnnf{\clist_{\ell}}$}
\DisplayProof
\end{center}
For each $1 \leq i \leq n$, we take $\seq_{i}'$ and each sequent proven as shown above, apply the $\conr$ rule $n{-}1$ times, and then $\wk$ to obtain a proof of the sequent $\seq_{i}$ as shown in the proof below. 
\begin{center}
\AxiomC{$\rel', \Psi, \neqset{y_{0}, \ldots, y_{n}}, \rel'' \sar \cxtii', \cxtiii', y_{i} : \bigsqcap \negnnf{\clist_{\ell}} \ | \ 1 \leq \ell \leq n$}
\RightLabel{$\conr \times n{-}1$}
\UnaryInfC{$\rel', \Psi, \neqset{y_{0}, \ldots, y_{n}}, \rel'' \sar \cxtii', \cxtiii', y_{i} : \bigsqcap \negnnf{\clist}$}
\RightLabel{$\wk$}
\UnaryInfC{$\rel', \Psi, \neqset{y_{0}, \ldots, y_{n}}, \rel'' \sar \cxtii', \cxtiii',x : \geqnir. \negnnf{\bigsqcup \clist}, y_{i} : \negnnf{\bigsqcup \clist}$}
\DisplayProof
\end{center}
Note that the last inference inference above is warranted since $\bigsqcap \negnnf{\clist} = \negnnf{\bigsqcup \clist}$ by definition, and because $\wk$ is hp-admissible by \lem~\ref{prop:hp-admiss-wk}. Next, observe that for each $i$ and $j$ such that $0 \leq i < j \leq n$, the following sequent $\seq_{i,j}$ is provable by $\ideq$.
\begin{center}
\AxiomC{}
\RightLabel{$\ideq$}
\UnaryInfC{$\rel', \Psi, \neqset{y_{0}, \ldots, y_{n}}, \rel'', y_{i} \eq y_{j} \sar \cxtii', \cxtiii', x : \geqnir. \negnnf{\bigsqcup \clist}$}
\DisplayProof
\end{center}
We can finish the proof of claim (2) as follows:
\begin{center}
\AxiomC{$\seq_{0} \ \cdots \ \seq_{n} \ \ \seq_{0,1} \ \cdots \ \seq_{n,n+1}$}
\RightLabel{$(\geqnir)$}

\UnaryInfC{$\rel', \Psi, \neqset{y_{0}, \ldots, y_{n}}, \rel'' \sar \cxtii', \cxtiii', x : \geqnir. \negnnf{\bigsqcup \clist}$}
\RightLabel{$\ltnr$}
\UnaryInfC{$\rel, \Psi, \rel'' \sar \la : {\leqslant} n \role . C, \cxtii, \cxtiii', x : \geqnir. \negnnf{\bigsqcup \clist}$}
\DisplayProof
\end{center}
This completes the proof of the lemma.
\end{proof}

\begin{customlem}{\ref{lem:sub-interpolation-lemma-3}}
If $\rel \sar \cxti, \cxtii$ is provable in $\calc$ for all $(\ \vdash \cxtii) \in \itp$ and $\lab(\itp) = \{\la\}$, then $\calc \Vdash \rel \sar \cxti, \la : \bigsqcap \bigsqcup \itp$. 
\end{customlem}

\begin{proof} Let $\itp := \{( \ \sar x : C_{i,1}, \ldots, x : C_{i,k_{i}}) \ | \ 1 \leq i \leq n\}$ and suppose that $\rel \sar \cxti, x : C_{i,1}, \ldots, x : C_{i,k_{i}}$ is provable in $\calc$ for $1 \leq i \leq n$. By repeated application of the $\disr$ rule, we obtain $\rel \sar \cxti, x: \bigsqcup_{1 \leq j \leq k_{i}} C_{i,j}$ for each $1 \leq i \leq n$. Applying the $\conr$ rule $n{-}1$ times lets us prove $\rel \sar \cxti, x: \bigsqcap_{1 \leq i \leq n} \bigsqcup_{1 \leq j \leq k_{i}} C_{i,j}$, which gives us the desired conclusion. 
\end{proof}

\begin{customlem}{\ref{lem:sub-interpolation-lemma-4}}
If $\rel \sar \cxti, \cxtii$ is provable in $\calc$ for all $(\ \sar \cxtii) \in \orth{\itp}$ and $\lab(\itp) = \{\la\}$, then $\calc \Vdash \rel \sar \cxti, \la : \negnnf{\bigsqcap \bigsqcup \itp}$. 
\end{customlem}

\begin{proof} Let $\itp = \{( \ \sar \cxtii_{1}), \ldots, ( \ \sar \cxtii_{n})\}$ such that $\cxtii_i = \{x: C_{i,1}, \ldots, x: C_{i,k_i} \}$ for each $1 \leq i \leq n$. We show $\rel \sar \cxti, \la : \negnnf{\bigsqcap \bigsqcup \itp}$ is provable in $\calc$ by induction on the cardinality of $\itp$.\\ 

\noindent
\textit{Base case}. 
 Let us suppose that $\itp$ is a singleton. Then, by assumption $\rel \sar \cxti, x : \neg C_{1,j}$ is provable for all $1 \leq j \leq k_{1}$, and so, the conclusion follows by $k_1 {-} 1$ applications of $\conr$.\\

\noindent
\textit{Inductive Step}. Suppose that $\itp = \{(\ \sar \cxtii_{1}), \ldots, (\ \sar \cxtii_{n+1})\}$ contains $n+1$ elements and assume that $\rel \sar \cxti, \cxtiii$ is provable in $\calc$ for all $\cxtiii \in \orth{\itp}$. Then, for each $1 \leq j \leq k_{n+1}$, the sequent $\rel \sar \cxti, x : \negnnf{C}_{n+1,j}, \cxtii'$ is provable in $\calc$ for each $( \ \sar \cxtii') \in  \orth{(\itp \setminus \{( \ \sar \cxtii_{n+1})\})}$. By IH, for each $1 \leq j \leq k_{n+1}$, the following sequent is provable in $\calc$.
$$
\rel \sar \cxti, x : \bigsqcup\bigsqcap \negnnf(\itp \setminus \{(\ \sar \cxtii_{n+1})\}), x: \negnnf{C}_{n+1,j}
$$
By applying the $\conr$ rule $k_{n+1} {-} 1$ times between each of the sequents above, followed by a single application of the $(\sqcup)$ rule, we obtain a proof of
$$
\rel \sar \cxti, x : \bigsqcup\bigsqcap \negnnf(\itp \setminus \{(\ \sar \cxtii_{n+1})\}) \sqcup \!\!\!\! \bigsqcap_{1\leq j \leq k_{n+1}} \!\!\!\! \negnnf{C}_{n+1,j},
$$
which gives our desired conclusion because
$$
\negnnf{\bigsqcap \bigsqcup \itp} = \bigsqcup\bigsqcap \negnnf(\itp \setminus \{(\ \sar \cxtii_{n+1})\}) \sqcup \!\!\!\! \bigsqcap_{1\leq j \leq k_{n+1}} \!\!\!\! \negnnf{C}_{n+1,j}.\qedhere
$$
\end{proof}

\begin{customthm}{\ref{thm:interpolation}}
Let $\ont = \ont_{1} \cup \ont_{2}$ be a $\riq$ ontology. If $\ont \vDash C \imp D$, i.e. $\calc, \prf \Vdash C \imp D$, then a concept interpolant $I$ can be computed in $\exptime$ relative to $\size{\prf}$ such that $\calc \Vdash C \imp I$ and $\calc \Vdash I \imp D$, i.e. $\ont \vDash C \imp I$ and $\ont \vDash I \imp D$.
\end{customthm}

\begin{proof} By our assumption and \cor~\ref{cor:subsumption-sequent-equiv}, it follows that $\calc, \prf \Vdash C \imp D$, meaning $\sar x : \gcilist{\ont}, x : \negnnf{C} \dis D$ is the conclusion of $\prf$ by definition. By \lem~\ref{prop:hp-invert}, namely, the hp-invertibility of the $\disr$ rule, $\sar x : \gcilist{\ont}, x : \negnnf{C}, x : D$ is provable in $\calc$ with a proof $\prf'$. By \lem~\ref{lem:main-interpolation-lemma-1}, the interpolation sequent $\emptyset; \emptyset \splti \emptyset \sar x : \gcilist{\ont_{1}}, x : \negnnf{C} \splti x : D, x : \gcilist{\ont_{2}} \sep \itp$ is provable in $\icalc$ with $\lab(\itp) = \{x\}$ and $\cpt{\itp} \subseteq \cpt{\ont_{1}, \negnnf{C}} \cap \cpt{D, \ont_{2}}$. Moreover, as this interpolation sequent is free of (in)equalities, i.e. $\rel,\Phi, \Psi = \emptyset$, we note that by the first two properties of \lem~\ref{lem:main-interpolation-lemma-1}, $\itp$ must be of the following form:
$$
\itp = \{( \ \sar x : C_{i,1}, \ldots, x : C_{i,k_{i}}) \ | \ 1 \leq i \leq n\}.
$$
 By Lemmas~\ref{lem:interpolation-lemma} -- \ref{lem:sub-interpolation-lemma-4}, we have $\sar x : \gcilist{\ont_{1}}, x : \negnnf{C}, x : I$ and $\sar x : \gcilist{\ont_{2}}, x : D, x : \negnnf{I}$ are provable in $\calc$ such that $I = \bigsqcap \bigsqcup \itp$. Both of the sequents $\sar x : \gcilist{\ont_{1}}, x : \negnnf{C} \dis I$ and $\sar x : \gcilist{\ont_{2}}, x : \negnnf{I} \dis D$ are provable with proofs $\prf_{1}$ and $\prf_{2}$ in $\calc$ by applying $\disr$, respectively. By \cor~\ref{cor:subsumption-sequent-equiv}, $\ont \vDash C \imp I$ and $\ont \vDash I \imp D$.

Last, we argue that the computation of $I$ is in $\exptime$ relative to the size of $\prf$. Note that applying \lem~\ref{lem:main-interpolation-lemma-1} is composed of the following steps: first, we partition the sequents in $\prf'$ by starting with the conclusion of $\prf'$ and working our way up the proof toward the initial sequents placing active formulae of rule instances within the same component of the partition as their corresponding principal formulae. Second, we assign interpolants to all of the initial sequents and work our way back down toward the conclusion of the proof by computing the interpolant of the conclusion of a rule from the interpolants of its premises, and applying the $\orthru$ when needed. The only problematic operation is the $\orthru$ rule as it may exponentially increase the size of interpolants, meaning the calculation of $I$ may be exponential in the worst-case.
\end{proof}